\RequirePackage[2020-02-02]{latexrelease}
\documentclass[prl,preprintnumbers,twocolumn,aps,floatfix]{revtex4}
\usepackage{graphicx,soul}
\usepackage{graphicx,dcolumn,bm,epsfig,amsmath,amssymb,textcomp,}
\usepackage{float}
\usepackage[utf8x]{inputenc}

\usepackage{xcolor}
\usepackage{bbm}

\begin{document}

\title{Effects of magnetic field on the evolution of energy 
density fluctuations}

\author{Shreyansh S. Dave\footnote{shreyansh.dave@tifr.res.in}}
\author{Subrata Pal\footnote{spal@tifr.res.in}}
\affiliation{Department of Nuclear and Atomic Physics, 
Tata Institute of Fundamental Research, Mumbai 400005, India}

\begin{abstract}
We study the effects of a static and uniform magnetic field
on the evolution of energy density fluctuations present in a
medium. By numerically solving the relativistic Boltzmann-Vlasov
equation within the relaxation time approximation, we explicitly
show that magnetic field can affect the characteristics of energy
density fluctuations at the timescale the system achieves local
thermodynamic equilibrium. A detailed momentum mode analysis of
fluctuations reveals that magnetic field increases the damping of
mode oscillations, especially for the low momentum modes. This
leads to a reduction in the ultraviolet (high momentum) cutoff of
fluctuations and also slows down the dissipation of relatively low
momentum fluctuation modes. We discuss the phenomenological
implications of our study on various sources of fluctuations in
relativistic heavy-ion collisions.
\end{abstract}

\maketitle

\noindent 

\section*{I. Introduction}

The effects of magnetic field on the bulk evolution 
of dynamical systems have been extensively studied and 
found to be crucial for a proper understanding of the  
physical properties of the system. For example, the
evolution of cosmic fluid in the early Universe has 
been found to be affected by a primordial magnetic 
field that can non-trivially modify the power spectrum 
of cosmic microwave background radiation (CMBR) 
\cite{cmbr_mag}. The fluid evolution in the post-merger 
(ringdown) phase of a binary neutron star merger and
the gravitational collapse of a homogeneous dust 
\cite{cllpsmag} can be influenced by the magnetic field
\cite{BNSmag0,BNSmag1,BNSmag2,BNSmag3,BNSmag4} modifying 
the strain amplitude and frequency spectrum of the
gravitational waves \cite{BNSmag5,BNSmag6,BNSmag7,
BNSmag8}. A strong magnetic field can also be generated
in the participant zone in relativistic heavy-ion
collisions \cite{larry,lateTmB} which can qualitatively modify
the azimuthal anisotropic flow and power spectrum of the
flow fluctuations of the hadrons \cite{ranjita,tuchin0,KharzMHD,
ourMHD,ourRev,mhd1,mhd2}.

These systems were investigated within the relativistic
magnetohydrodynamic (RMHD) framework where the medium is
assumed to be in the local thermodynamic equilibrium. The
observable consequences of magnetic field basically stem
from the stiffening of the equation of state, which causes
an increase in the sound speed in the plane perpendicular
to the magnetic field and generates an additional momentum
anisotropy in the fluid evolution \cite{Landau_em,ranjita,
ourMHD,ourRev}.

However, expanding systems are naturally not in thermal
equilibrium, but may gradually approach equilibrium from
out-of-equilibrium initial conditions. Such scenarios
have been perceived in the reheating of early Universe
in the presence of a magnetic field \cite{magInfl1,
magInfl2}, and in noncentral relativistic heavy-ion
collisions where a deconfined state of Quark-Gluon-Plasma
(QGP) \cite{RevQGP1,RevQGP2,RevQGP3} is formed in presence
of a large magnetic field \cite{larry,lateTmB,
tuchin1,tuchin2,BMEq}. In these situations, the RMHD
framework is not applicable and an out-of-equilibrium
description is required to explore the evolution of the
physical quantities as they approach local-equilibrium.

In general, fluctuations in physical quantities can
exist at various length scales and may largely influence
the dynamics of the system. In fact, fluctuations can
be exploited to infer the information of a system at
various time/length scales. For cosmic fluid and
astrophysical systems, compared to long wavelength
fluctuations, the short wavelength fluctuations (comparable
to the coarse-grained length scale for hydrodynamic 
description) may not have any significant effect on the
bulk evolution.  In contrast, in heavy-ion collisions,
fluctuations of short wavelengths comparable to the length
scale, $\ell \sim 1$ fm, are
particularly important in the description of the evolution
dynamics of QGP droplet of transverse length $L \sim 10$
fm. These fluctuations are dominantly present at the
initial stages of the collision \cite{hydroSim,Schenke12},
and also during the space-time evolution (collisional and
thermal/hydrodynamic fluctuations), and play an important
role in the bulk hydrodynamic description of the QGP medium
\cite{kapusta1,spal1,stephanov1}.$^1$\footnotetext[1]{
Additionally, at the moderate collision energies, large
critical fluctuations can inevitably arise near the QCD
critical point and can potentially be used to probe the
critical point \cite{stephanov1,stephanov2}.}

While these model studies of fluctuations were carried
out in the absence of magnetic field, the short-lived
strong magnetic field produced in nuclear collisions
\cite{larry} can affect mainly the ``fast-evolving"
short wavelength fluctuations and thereby the observables
that are sensitive to fluctuations. Thus, it is important
to study the impact of magnetic field on these
fluctuations that can provide reliable insight on the
medium properties in a model-to-data comparison
\cite{ourRev,saumia1,Schenke12i,saumia2,ehr}.$^2
$\footnotetext[2]{Interestingly, the ideal RMHD
simulations of QGP in relativistic heavy-ion collisions
mostly show the effect of magnetic field on the higher
flow harmonics \cite{ourMHD}.}

In this work, we investigate within kinetic transport
\cite{rischke1,rischke2,Ashu1}, the effects of an
external magnetic field on the evolution of energy
density fluctuations present in a slightly out-of-equilibrium
medium of electrically charged particles. The magnetic
field $B$ is considered to be static, uniform, and
marginally weaker than the thermal energy or the
temperature $T$ of the medium, i.e., $\sqrt{|q B|} <
T$. In particular, we solve numerically the relativistic
Boltzmann-Vlasov equation within the relaxation time
approximation (RTA) \cite{bgk,Anderson,PaulR}, and show
that the magnetic field can affect the evolution of
energy density fluctuations in the transverse direction
to $B$, leading to fluctuations of completely different
characteristics at the timescale at which the medium
achieves local-equilibrium. We perform momentum mode
analysis of the fluctuations and demonstrate that the
magnetic field enhances the damping of mode oscillations. We
extend the analysis to the high momentum scale of
fluctuations where the nonhydrodynamic modes of RTA
kinetic theory dominate \cite{PaulR0,heinz1,PaulR,
kurkela1,PaulR1,kurkela2,ehr} and determine the ultraviolet
cutoff of fluctuations above which all the higher momentum
modes get suppressed while approaching local-equilibrium. We
show that this cutoff decreases (i.e., the short-wavelength
cutoff increases) with increasing magnetic field.

We emphasize that the analysis presented here is quite
general and can be applied to any system whose constituents 
are electrically charged. For inclusiveness, we consider a 
system of charged pions and discuss the phenomenological
implications on relativistic heavy-ion collisions.

This paper is organized as follows. Section II deals with 
a detailed description of the Boltzmann-Vlasov equation, 
where the underlying assumptions for solving this equation
are discussed. The simulation details are given in Sec.
III, and the simulation results are presented in Sec. IV.
In Sec. IVA, the effects of magnetic field on the evolution
of energy density fluctuations are studied with an extensive
analysis of the momentum modes of fluctuations. The effects
of magnetic field on the other components of energy-momentum
tensor are shown in Sec. IVB. In Sec. IVC, the effects of
magnetic field on a generic initial energy density profile
are presented, which readily elucidate the implications of
the preceding sections. The phenomenological implications of
the results, especially in the context of QGP formation in
relativistic heavy ion collisions are discussed in Sec. V.
Finally, we conclude with a summary of the work in Sec. VI.

Throughout the study, we consider the Minkowski space-time
metric as $\eta_{\mu \nu}= {\rm diag}(1,-1,-1,-1)$, and work
in the units $k_B=\hbar=c=1$. The four-position and
four-momentum of particle (and anti-particle) are
represented by $x^\mu=(t,{\bf x})$ and $k^\mu=(k^0,{\bf k})$,
respectively, where $k^\mu$ is normalized to the particle's
rest mass square as $k^\mu k_\mu=m_0^2$ giving $k^0=
\sqrt{{\bf k}^2+m_0^2}$ =$E_{_{\bf k}}$ \textemdash$
~$the energy of particle with three-momentum ${\bf k}$.

\section*{II. Boltzmann-Vlasov Equation}

To study the effects of magnetic field on the equilibration 
of a system, we solve the relativistic Boltzmann-Vlasov (BV)
equation \cite{groot,rischke2}:
\begin{equation}
  k^\mu \partial_\mu f + q F^{\mu \nu} k_\nu 
 \frac{\partial}{\partial k^\mu} f = \mathcal{C}[f,\bar{f}].
\label{BV:eq}
\end{equation}
This provides the time evolution of the single-particle
phase-space distribution function $f\equiv f(t,{\bf x},
{\bf k})$ of particles with electric charge $q$. Here $F^{\mu
\nu}$ is the electromagnetic field tensor whose components
are treated as external fields.$^3$\footnotetext[3]{Thus, unlike
a RMHD fluid \cite{ourMHD}, there is no feedback of the
medium on the electromagnetic fields.} The collision integral
$\mathcal{C}[f,\bar{f}]$ makes the BV equation nonlinear
which cannot be solved analytically \cite{bamps1,transMag1,
BMEq}. We consider the linear approximation \textemdash$
~$also known as the relaxation time approximation (RTA)
\cite{bgk,Anderson,PaulR} \textemdash$~$where the system is
assumed to be slightly away from the equilibrium state such
that the distribution function can be written as $f=f_{\rm eq}
+\delta f$, where $f_{\rm eq}$ is the local equilibrium
distribution function of the system and $\delta f \ll 
f_{\rm eq}$ gives the deviation from $f_{\rm eq}$.

In RTA, the collision integral can be written in the linear
form as $\mathcal{C}[f,\bar{f}]$=$-\frac{k^\mu u_\mu}{\tau_c}
\delta f$ \cite{Anderson,Ashu1}, where $\tau_c$ is the
relaxation time which sets a timescale for local equilibration
\cite{Anderson,PaulR}, $u^\mu$=$\gamma (1,{\bf v})$ the
four-velocity of the fluid, and $\gamma$=$(1-{\bf v}^2)^{-1/2}$
the Lorentz factor. For anti-particles, the Boltzmann-Vlasov
equation has a form similar to Eq. \eqref{BV:eq} with $(f,q)
\leftrightarrow (\bar{f},-q)$ and the collision term under
the RTA as $\bar{\mathcal{C}}[f,\bar{f}]$=$-\frac{k^\mu 
u_\mu}{\tau_c} \delta \bar{f}$.$^4$\footnotetext[4]{Note that
a magnetic field can modify the relaxation time $\tau_c$ 
\cite{transMag2}. However, we have rescaled the space-time
coordinates by $\tau_c$ and hence it will not enter
in the BV equation explicitly.}
The collision term within RTA is constrained by the Landau
matching conditions required to satisfy the net-particle
four-current and energy-momentum conservations \cite{landau_k,
Anderson,sunil2022}. These conditions are given by
\begin{equation}
\begin{split}
u_\mu T^{\mu \nu} = u_\mu T^{\mu \nu}_{\rm eq},\\
u_\mu N^\mu = u_\mu N^\mu_{\rm eq},
\end{split}
\label{matchEq}
\end{equation}
which should be satisfied throughout the evolution of
distribution functions \cite{Anderson,sunil2022}. Here
$T^{\mu \nu}$ and $T^{\mu \nu}_{\rm eq}$ are the energy-momentum
tensors of the medium corresponding to distribution
functions ($f,\bar{f}$) and local equilibrium distribution
functions ($f_{\rm eq},\bar{f}_{\rm eq}$), respectively.
Likewise, $N^\mu$ and $N^\mu_{\rm eq}$ are the net-particle
four-currents corresponding to ($f,\bar{f}$) and ($f_{\rm eq},
\bar{f}_{\rm eq}$), respectively. These variables can
be calculated by using the relations \cite{groot,sunil2022}
\begin{equation}
\begin{split}
 T^{\mu \nu} = \int dK ~ k^\mu k^\nu (f+\bar{f}),\\
 N^\mu = \int dK ~ k^\mu (f-\bar{f}),~~~
\end{split}
\label{eq:tmn}
\end{equation}
where $dK=d^3 k/[(2\pi)^3E_{_{\bf k}}]$ is the Lorentz
invariant momentum space integration measure. The Landau
matching conditions are satisfied by the Landau-Lifshitz's
definition of four-velocity $u^\mu$ \cite{sunil2022,groot}:
\begin{equation}
u^{\mu}=\frac{T^{\mu\nu}u_{\nu}}{u_\rho T^{\rho \sigma} 
u_{\sigma}}.
\end{equation}
In this definition, the momentum density and energy flux
are zero in the local rest frame of the medium.$^5
$\footnotetext[5]{The Eckart's definition of fluid velocity
\cite{groot} becomes ambiguous at zero chemical potential and
therefore not used here.}

We consider a static and uniform magnetic field along
$y$ direction, i.e., ${\bf B}$=$B_0\hat{y}$, which yields
the BV equation of the form
\begin{align} \label{BVsim:eq}
 & E_{_{\bf k}} \frac{\partial f}{\partial t} + 
 k_x \frac{\partial f}{\partial x} +
 k_y \frac{\partial f}{\partial y} + 
 k_z \frac{\partial f}{\partial z}
 + q B_0 \Big(k_x \frac{\partial f}{\partial k_z} -
 k_z \frac{\partial f}{\partial k_x} \Big) \nonumber\\
 & = \frac{k^\mu u_\mu}{\tau_c} (f_{\rm eq} - f).
\end{align}
The magnetic field thus affects the distribution function
in the $(k_x,k_z)$ plane, but has no effect in the $k_y$
direction. This suggest that during evolution, the magnetic
field can by itself generate three-dimensional spatial
anisotropies of any fluctuation present in the system.
Further, in the RTA (without non-linearity), direct coupling
between the fluctuation modes in the transverse ($xz$) plane
and the parallel ($y$) direction is not expected. Consequently,
such anisotropies can grow in the linear regime until the
fluctuations decay to vanishingly small values.

Solving the integro-differential Boltzmann-Vlasov equation
in (6+1)-dimensional phase-space becomes computationally
quite intensive as we are interested in studying the
short-wavelength fluctuations that require small lattice
spacing and hence large number of lattice points. We take
recourse to a tractable (3+1)-dimension equation, where we
consider the evolution of distribution function $f \equiv
f(t,x,k_x,k_z)$ for magnetic field pointing along $y$
direction. This implies a variation of $f$ in phase-space
along $x$ direction and homogeneity along $y$ direction with
$k_y=0$.$^6$\footnotetext[6]{This assumption will not alter
our final conclusions as the evolution of the distribution
functions ($f, \bar{f}$) is unaffected by the magnetic field
along $y$ direction.} Further, $f$ is taken homogeneous
along spatial $z$ direction, but its variation is accounted
along $k_z$. This allows us to study the effect of magnetic
field that generates finite values of $T^{0z}$ and $T^{xz}$.$
^7$\footnotetext[7]{We have checked the reliability of our
results presented in Sec. IV by performing a (4+1)D
phase-space simulation, namely with $f(t,x,z,k_x,k_z)$ but
using larger lattice spacing, which then simulates {\it
long-wavelength fluctuations}. We find that the qualitative
aspects of the results remain unchanged with the inclusion of
inhomogeneity along spatial $z$ direction.
The aspects of long-wavelength (hydrodynamic)
fluctuations in the (4+1)D code will be presented in a separate
communication.} For numerical
simulations we rewrite the BV equations for particles and
anti-particles in the dimensionless form 
\begin{equation}
\begin{split}
 E_{_{\bf k'}}\frac{\partial f}{\partial t'} + 
 k'_x \frac{\partial f}{\partial x'} + 
 \beta_0 \Big(k_x' \frac{\partial f}{\partial k_z'} -
 k_z' \frac{\partial f}{\partial k_x'} \Big) = 
 k'^\mu u_\mu (f_{\rm eq} - f),
 \\
 E_{_{\bf k'}}\frac{\partial \bar{f}}{\partial t'} + 
 k'_x \frac{\partial \bar{f}}{\partial x'} - 
 \beta_0 \Big(k_x' \frac{\partial \bar{f}}{\partial k_z'} -
 k_z' \frac{\partial \bar{f}}{\partial k_x'} \Big) = 
 k'^\mu u_\mu (\bar{f}_{\rm eq} - \bar{f}),
 \end{split}
 \label{eomF}
\end{equation}
where we have introduced the dimensionless variables 
$t'$=$t/\tau_c$, $x'$=$x/\tau_c$, ${\bf k'}$=$
{\bf k}/m_0$ (hence $E_{_{\bf k'}}$=$E_{_{\bf k}} /m_0$), 
with $|{\bf k}|=\sqrt{k_x^2+k_z^2}$. The term $\beta_0
$=$|qB| \tau_c/m_0$, is dubbed as the \textquotedblleft
magnetic field parameter" which is varied to study the
impact of magnetic field $B$ on the medium evolution.
The two coupled differential equations in Eq.
\eqref{eomF} are simultaneously solved numerically for
complete dynamical evolution. To ensure energy-momentum
and net-particle four-current conservations by the RTA
collision kernel, the Landau matching conditions given
in Eq. \eqref{matchEq} are imposed, i.e., $\varepsilon =
\varepsilon_{\rm eq}(T,\mu)$ and $n = n_{\rm eq}(T,\mu)$
\cite{Anderson,sunil2022,spal2,
chandro22}. The energy density
$\varepsilon = u_\mu u_\nu T^{\mu \nu}$ and the net-particle
density $n = u_\mu N^\mu$ can be obtained from Eq.
\eqref{eq:tmn}. Similarly, the equilibrium energy density
$\varepsilon_{\rm eq} = u_\mu u_\nu T^{\mu \nu}_{\rm eq}$
and the net-particle density $n = u_\mu N^\mu_{\rm eq}$
can be obtained from Eq. \eqref{eq:tmn} by using the
local-equilibrium distributions
\begin{equation}
 f_{\rm eq} = \frac{1}{\exp{(\beta E_{_{\bf k}} - \alpha)}
 \pm 1},~
 \bar{f}_{\rm eq} = \frac{1}{\exp{(\beta E_{_{\bf k}} +
 \alpha)} \pm 1},
\end{equation}
where $\pm$ refer to fermions and bosons, respectively.
The associated equilibrium temperature $T=1/\beta$,
chemical potential $\mu=\alpha/\beta$ are solved by using
the matching conditions. It may be noted that
alternative collision kernels are also available,
such as the Bhatnagar-Gross-Krook (BGK) collision kernel
\cite{bgk} and its relativistic extensions \cite{bgk1,bgk2,rafelski1,rafelski2,
pracheta} that conserves
net-particle four-current, and the recently proposed novel RTA
\cite{novelRTA}.

To generate the initial configuration of $f$ (and 
$\bar{f}$), the following procedure is adopted. We
start with a local equilibrium state defined by a 
temperature $T(x')$ that gives a local equilibrium
distribution function $f_{\rm eq}$. Specifically, we
consider sine-Gaussian function for temperature
fluctuations which gives
\begin{equation}
 T(x')=T_0 + \delta T \sin\Big(\frac{2\pi r x'}{L}\Big)
 \exp (-x'^2/2 \chi^2),
 \label{eq:fluct}
\end{equation}
where $T_0$ resembles the global equilibrium temperature
with a corresponding distribution function $f_0$.
$\delta T$ is the temperature fluctuation scale
factor which is taken sufficiently small $\delta T$=0.01$
T_0$. This ensures that long timescales are required to
smooth out the inhomogeneities in $T(x')$ and achieve
global equilibrium in the system of total size $L$. The
Gaussian width is taken to be $\chi$=$L/10$ and the
\textquotedblleft fluctuation parameter" $r$ is varied to
simulate the effects of magnetic field on different
wavelength fluctuations. This procedure also ensures that
the spatial variations in $f_{\rm eq}$ are sufficiently
localized compared to the total system size $L$.

We perturb further the local equilibrium state in the 
($x',k'_x,k'_z$) space such that the system achieves an
out-of-equilibrium state. For this purpose, we consider
random fluctuations, $\delta f$, on top of $f_{\rm eq}$,
which gives the initial distribution function as $f$=$
f_{\rm eq}$+$\delta f$. The choice of $\delta f$ is
dictated by the Landau matching conditions \cite{landau_k,
Anderson} such that the initial out-of-equilibrium
distribution function, $f$, gives the same net-particle
number density and energy density as that given by $
f_{\rm eq}$. The random fluctuation $\delta f$ is then
taken as
\begin{equation}
\delta f = \mathcal{R}_{x'k'_xk'_z} f_{\rm eq},
\label{eq:fluct0}
\end{equation}
where $\mathcal{R}_{x'k'_xk'_z}$ is a small random number
in the ($x',k'_x,k'_z$) space which varies from negative 
to positive values. The maximum value of $|\mathcal{R}_{x'
k'_xk'_z}|$ is set by the RTA condition $\delta f \ll
f_{\rm eq}$, which gives $|\mathcal{R}_{x'k'_x k'_z}|\ll 1$.
We set the maximum value of $|\mathcal{R}_{x'k'_xk'_z}| =
0.01$. Likewise, an initial configuration for $\bar{f}$ is
also generated. Note that our final conclusions of the study
are completely independent of the choices of initial
fluctuations.

The fluctuations given in Eqs. \eqref{eq:fluct} and
\eqref{eq:fluct0} may be considered as an individual
component out of all possible sources of non-equilibrium
fluctuations present in the system, such as the
initial-state or hydrodynamic fluctuations present in
relativistic energy nuclear collisions \cite{hydroSim,
Schenke12}$^8$\footnotetext[8]{For various modes of energy
density fluctuations in the transverse plane of the
colliding system, see also Ref. \cite{hicFluct}.}. It is
therefore important to analyze the effects of magnetic
field on the evolution of these fluctuations.

\section*{III. Simulation Details}

The Boltzmann-Vlasov equations \eqref{eomF} are solved by
performing numerical simulations on a lattice of dimension
$1500\times 500\times 500$ with 1500 lattice points along
$x'$ direction. The spatial and momentum step sizes are
chosen to be $\Delta x'$=$\Delta k_x'$=$\Delta k_z'$=0.01,
yielding a total system size of $L=15$ in spatial direction
and $2L_{_{k'}}=5$ in the momentum direction. The time step
for evolution of distribution functions is taken to be $
\Delta t'$=$\Delta x'/2$. At the initial time, the medium
velocity is considered to be zero. The simulations are
performed using the second-order Leapfrog method
\cite{leapfrog} with periodic boundary conditions along all
the three directions (one space and two momentum).

With the given choice of simulation parameters, it is
convenient to consider a medium of charged pions $\pi^{\pm}$ 
(of mass $m_0 = m_\pi = 140$ MeV) which is slightly
away from the
equilibrium state with an ambient (global equilibrium)
temperature of $T_0=50$ MeV. The charge chemical potential
$\mu_Q$ is varied between $0-100$ MeV, though most of our
results are presented at $\mu_Q=100$ MeV.$^9
$\footnotetext[9]{At a non-zero $\mu_Q$, the effects of
magnetic field on the evolution of fluctuations become
explicit.} In the equilibrium, this pionic medium follows
the Bose-Einstein distribution function. 

In the study, the strength of magnetic field is
constrained by the thermal energy ($\sim T$) of the
medium \cite{rischke1} and taken in the marginally
weak field limit $\sqrt{|q B|} < T$ \cite{transMag1}. 
(A large $B$ would cause Landau quantization of the
energy levels in the transverse plane, which is not
the interest of the present study.) In terms of the
magnetic field parameter $\beta_0$ of Eq. \eqref{eomF},
this condition becomes equivalent to $\beta_0 < \tau_c
T^2/m_0$. For the pionic medium considered at $T \simeq
50$ MeV with $\mu_Q = 0-100$ MeV and for typical values
of relaxation time $\tau_c \gtrsim 15$ fm \cite{rlx1,rlx2},
the above condition gives $\beta_0 \lesssim 1$.
Accordingly, we have taken values of $\beta_0 = 0-0.7$
in this analysis.

It may be noted that the strength of the magnetic field
generated in the medium in relativistic heavy-ion
collision can be estimated from the difference in the
polarizations of $\Lambda$ and $\bar{\Lambda }$ hyperons
($\Delta \mathcal{P}=\mathcal{P}_\Lambda - \mathcal{P}_{
\bar{\Lambda}}$) via $B \propto T_f |\Delta \mathcal{P}
|$ \cite{lateTmB}. For STAR measurements of $\mathcal{P
}_\Lambda$ and $\mathcal{P}_{\bar{\Lambda}}$ in Au+Au
collisions at c.m. energy $\sqrt{s_{NN}} = 200$ GeV
\cite{star2}, a conservative upper bound of $|qB| < 2.7
\times 10^{-3} m_\pi^2$ was estimated at the level of one
standard deviation at a freeze-out temperature of $T_f =
150$ MeV. Since the observed value of $|\Delta \mathcal{P}
|$ is found to increase rapidly with decreasing collision
energy up to $\sqrt{s_{NN}} = 7.7$ GeV \cite{star1,star2,
kapusLambda}, the magnetic field $|qB|$ and thereby $\beta_0
$ could also increase (see also Ref. \cite{rafelski2}, where
the magnetic field was shown to increase at lower collision
energy mainly due to early freeze-out time).
Moreover, for three standard deviations, the upper bound on
the magnetic field strength becomes about six times larger
than the bound given above \cite{lateTmB}.

\section*{IV. Simulation Results and Discussions}

\subsection*{A. Energy density fluctuations}

In the Boltzmann-Vlasov simulation, we solve the
distribution functions of particles and anti-particles
at each space-time point and calculate the
\textquotedblleft dimensionless" energy density of the
system by using the relation
\begin{equation}
 \hat\varepsilon(x',t') = \int_{-L_{k'}}^{L_{k'}}
 \int_{-L_{k'}}^{L_{k'}} 
 d k_x' dk_z' ~ E_{_{\bf k'}} (f+\bar{f}).
 \label{eq:d2slv}
\end{equation}
The energy density fluctuations can be obtained from 
\begin{equation}
\delta \hat\varepsilon = \frac{1}{\hat\varepsilon_0} \{
\hat\varepsilon(x',t') - \langle \hat\varepsilon(x',t')\rangle \},
\end{equation}
where $\hat\varepsilon_0$ is the global equilibrium energy
density [calculated from Eq. \eqref{eq:d2slv} by using 
($f_0 , \bar{f}_0$)], and $\langle \hat\varepsilon(x',t')
\rangle$ is the average of $\hat\varepsilon (x',t')$ over
all lattice points at time $t'$. For our choice of $T_0
=50$ MeV and $\mu_Q=100$ MeV, one obtains $\hat\varepsilon_0
\approx 2.4$.

\begin{figure}[t]
\includegraphics[width=0.99\linewidth]{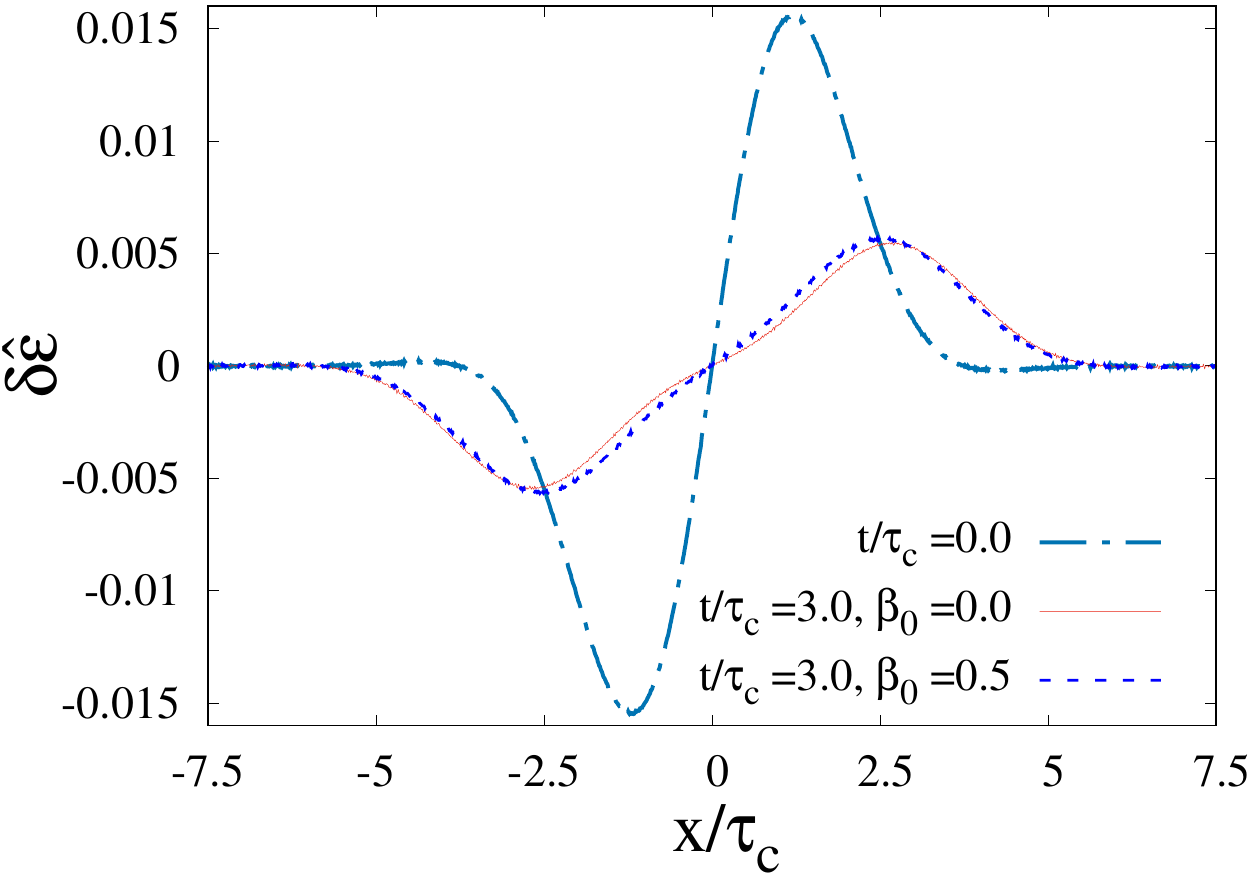}
 \caption{Energy density fluctuations at the initial time
 $t/\tau_c=0$ (dash-dotted line), and at time $t/\tau_c
 =3.0$ in the absence (solid line) and presence (dashed
 line) of magnetic field for the magnetic field parameter
 $\beta_0=0.5$ and the fluctuation parameter $r=2.0$.}
 \label{fig1}
\end{figure}

Figure \ref{fig1} shows the energy density fluctuations,
$\delta \hat\varepsilon$, at the initial time $t/\tau_c=0$
(dash-dotted line), and at time $t/\tau_c=3.0$ in the
absence (solid line) and presence (dashed line) of the
magnetic field, for the magnetic field parameter $\beta_0
= 0.5$ and the fluctuation parameter $r=2.0$. Compared to
the initial state, at later times the fluctuation spreads
out spatially and the peak amplitudes are dominantly
suppressed. The inclusion of magnetic field dampens the
evolution/expansion of the underlying medium, which in
turn slows down the propagation of the fluctuations
in the transverse direction. As a result, the fluctuations
persist with somewhat larger magnitudes and for longer
duration and should influence the final observables
compared to the magnetic-field-free situation.

\subsection*{1. Fourier modes of energy density fluctuations}

In this subsection, we analyze the effects of magnetic field
on energy density fluctuations in terms of momentum modes. We
note that in a non-equilibrium state various momentum modes
can be present, including very high momentum modes whose decay
timescales are lesser than or equal to the local-equilibration
timescale. Momentum of the critical mode (the mode whose decay
timescale is comparable to the local-equilibration timescale)
naturally sets an ultraviolet cutoff of fluctuations, above
which all higher momentum modes of fluctuations are suppressed
in local-equilibrium. In RTA, this cutoff is set by the local
equilibrium relaxation time $\tau_c$, where the modes with
momenta $\kappa \gtrsim \tau_c^{-1}$ are suppressed at the
timescale of $\tau_c$, while modes with momenta $\kappa
\lesssim \tau_c^{-1}$ can survive to further participate in
the (hydrodynamic or RMHD) evolution. In the following
analysis, we use this fact to set the momentum range
(wavelength) of initial fluctuations, by accordingly varying
the fluctuation parameter $r$ of Eq. \eqref{eq:fluct} in the
range $r \in [0.5,4.8]$.

\begin{figure}[t]
\includegraphics[width=0.99\linewidth, angle=180]{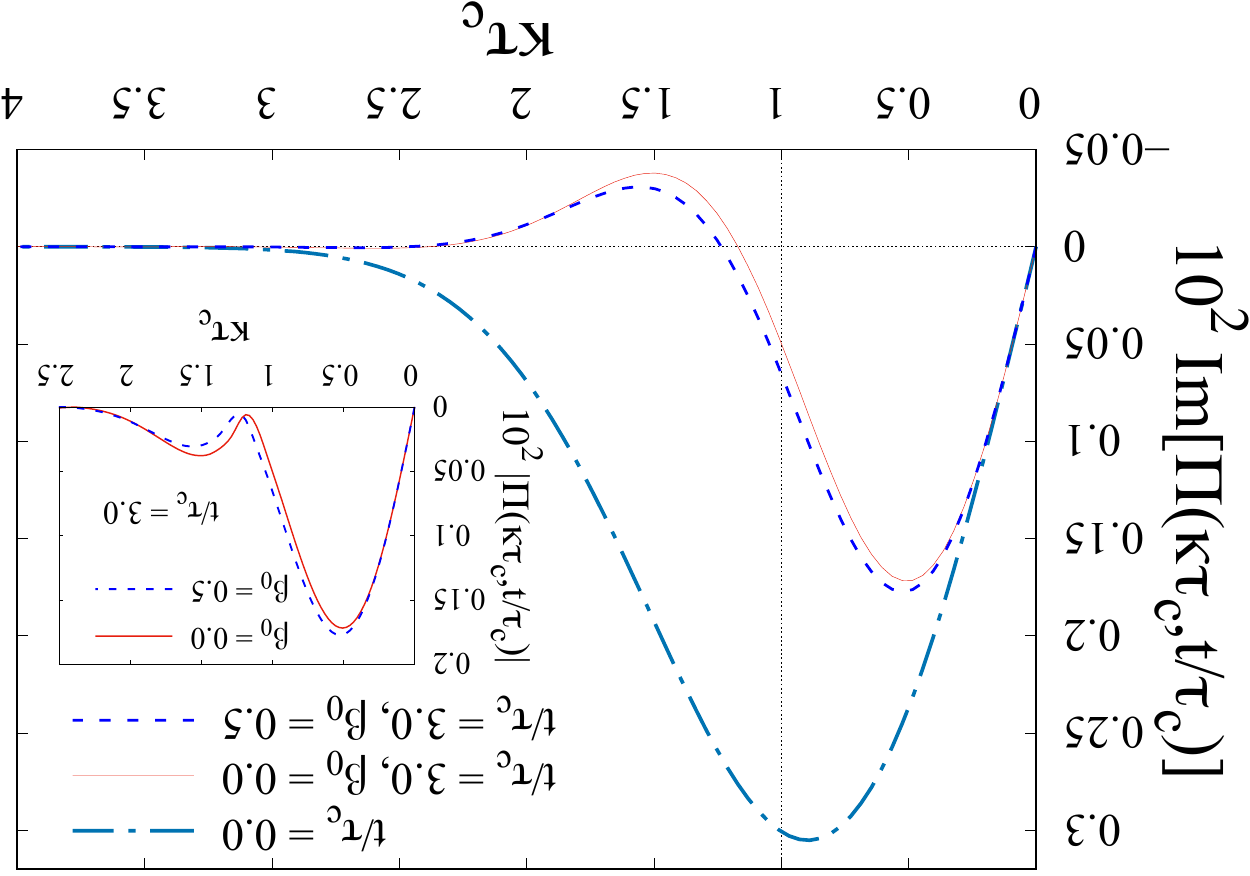}
 \caption{Momentum spectrum of energy density
 fluctuations at initial time (dash-dotted line), and at time
 $t/\tau_c = 3.0$ in the absence (solid line) and in the presence
 (dashed line) of magnetic field, for the same simulation
 parameters as used in Fig. \ref{fig1}. The dotted vertical line
 indicates the mode at momentum $\kappa \tau_c=1$. In the inset,
 the modulus of modes versus $\kappa \tau_c$ is plotted for $
 \beta_0=0$ and 0.5 at $t/\tau_c=3.0$.}
 \label{fig2}
\end{figure}

To quantify the evolution of energy density fluctuations 
with and without magnetic field, we perform Fourier
transform from configuration $x'$ space to the momentum
$\kappa'$ space as
\begin{equation}
  \Pi(\kappa',t') = \frac{1}{L}\int_{-L/2}^{L/2} dx'~
  \delta \hat\varepsilon(x',t')~e^{i \kappa' x'},
  \label{eq:four}
\end{equation}
where $\kappa' = \kappa \tau_c$ is the dimensionless momentum
of the mode $\Pi(\kappa',t')$, and $\kappa$ the dimensionful
momentum. Figure \ref{fig2} shows the momentum spectrum
(Im$[\Pi(\kappa',t')]$ versus $\kappa'$) of energy density
fluctuations at the initial time (dash-dotted line), and at
$t/\tau_c = 3.0$ in the absence (solid line) and in the presence
(dashed line) of magnetic field, for the same simulation
parameters as used in Fig. \ref{fig1}. At the initial time $t=0
$, the peak momentum of the spectrum at $\kappa_p \tau_c=0.88$
corresponds to the most dominant mode; the magnitude and position
of the peak depends on the fluctuation parameter $r$ of Eq.
\eqref{eq:fluct}. In the inset of Fig. \ref{fig2}, the modulus
of modes, $|\Pi(\kappa',t')|$, versus $\kappa \tau_c$ is plotted
for $\beta_0=0$ and 0.5 at $t/\tau_c=3.0$.

\begin{figure}[t]
\includegraphics[width=0.99\linewidth]{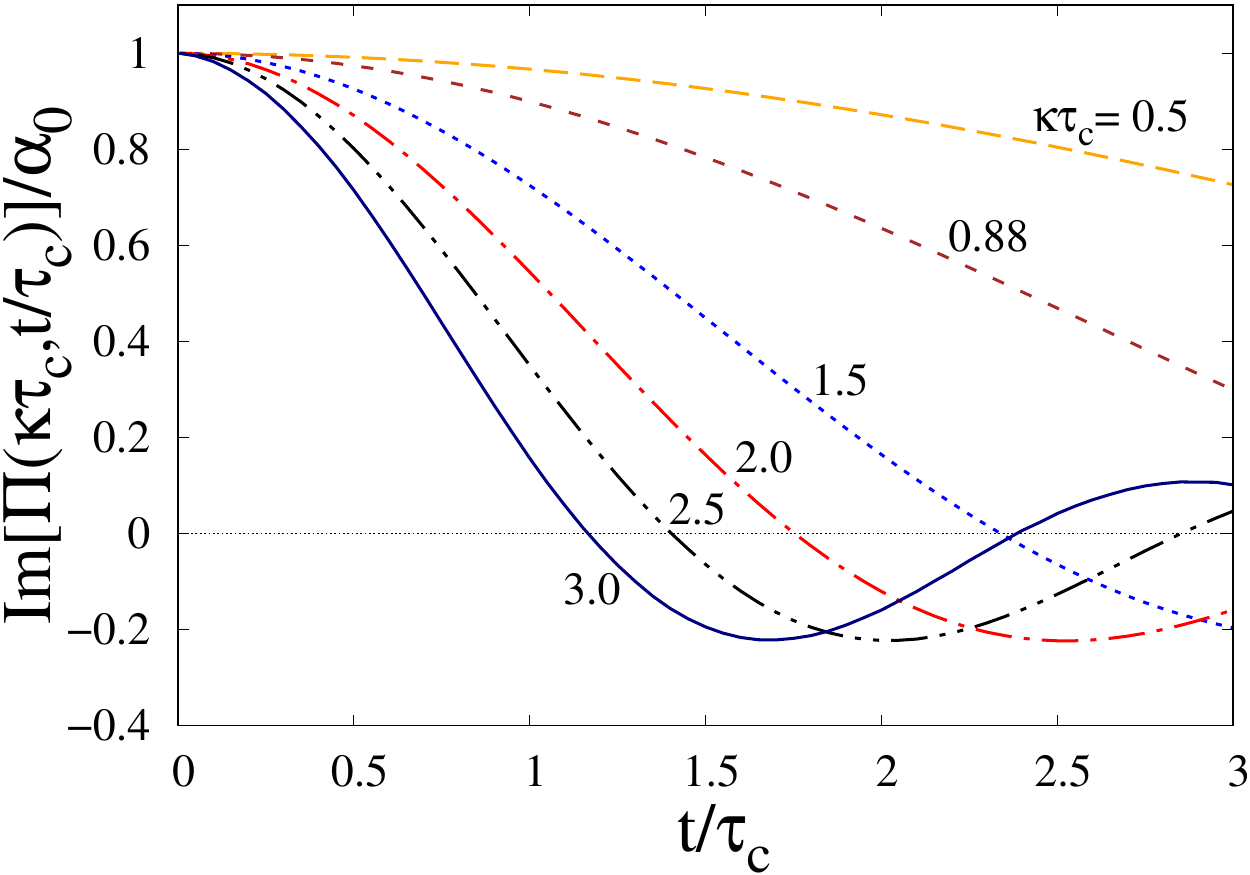}
 \caption{Damped harmonic oscillations of various non-zero modes
 present in the spectrum of Fig. \ref{fig2} (without magnetic field).
 Each mode is normalized with its initial value $\alpha_0$ (at $t/
 \tau_c=0$) and marked with corresponding momentum $\kappa \tau_c$.}
 \label{fig3}
\end{figure}

The modulus of modes, $|\Pi(\kappa',t')|$, quantify the strength
of fluctuations present in the system, thus as the fluctuations
evolve and damp, each mode Im$[\Pi(\kappa',t')]$ would decrease
with increasing time. Since the high momentum modes decay faster
compared to the low ones, the peak of the spectrum shifts towards
lower momenta at later times as can be seen at $t/\tau_c = 3.0$.
Moreover, certain higher momentum modes in the spectrum become
negative, which indicates that these modes are not just decaying
but also performing oscillations over time; see Fig. \ref{fig3}
for such damped harmonic oscillations of various modes present in
the spectrum.

In Fig. \ref{fig2} we also present the momentum spectrum in
presence of magnetic field at time $t/\tau_c=3.0$ (dashed
line). The magnetic field clearly affects the entire spectrum
of fluctuations, retaining to some extent the strength of the
initial fluctuations in the low momentum regime ($\kappa
\tau_c \lesssim 1.2$), while suppressing the {\it modulus} of
higher momentum modes relative to $B=0$ situation; see the
inset of Fig. \ref{fig2}. This leads to characteristic change in
the energy density fluctuations during the evolution towards
equilibrium. Later, we have shown that magnetic field increases
damping coefficient of mode oscillations causing such
characteristic changes in the fluctuations.

\begin{figure}[t]
\includegraphics[width=0.99\linewidth]{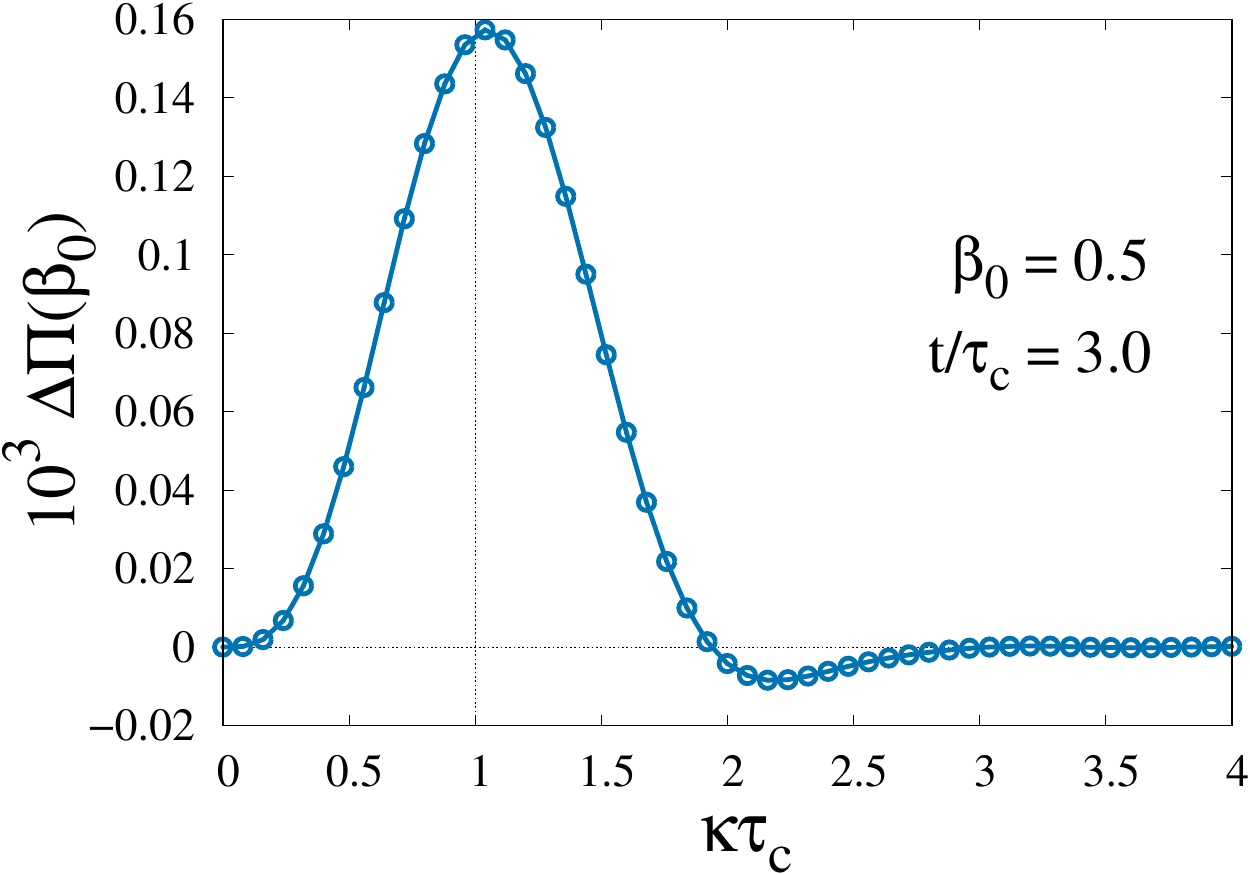}
 \caption{Momentum spectrum of difference of modes with and
 without magnetic field, $\Delta \Pi (\beta_0) = {\rm Im}[\Pi(\beta_0)]-
 {\rm Im}[\Pi(0)]$, at time $t/\tau_c = 3.0$ for the same
 simulation parameter as used in Fig. \ref{fig2}.}
 \label{fig4}
\end{figure}

To determine in which momentum regime the fluctuation modes are
strongly affected by the magnetic field at a given time, we
display in Fig. \ref{fig4} the momentum spectrum for the difference
of modes with and without magnetic field at time $t/\tau_c = 3.0$.
At the starting time of $t=0$, the momentum spectrum, with and
without $B$, are identical. Subsequently, at later times, the
magnetic field is found to influence various modes differently
leading to such a structure. The maximum difference, corresponding
to the peak position, occurs at a momentum $\kappa \tau_c \simeq
1.0$ which is larger than the peak-position momentum ($\kappa
\tau_c \simeq 0.5$) in Fig. \ref{fig2}. This implies that at any
instant, the magnetic field affects quantitatively more the
evolution of higher momentum modes of fluctuations as these fast
modes experience stronger Lorentz force (along $z'$ direction).

\begin{figure}[t]
\includegraphics[width=0.99\linewidth]{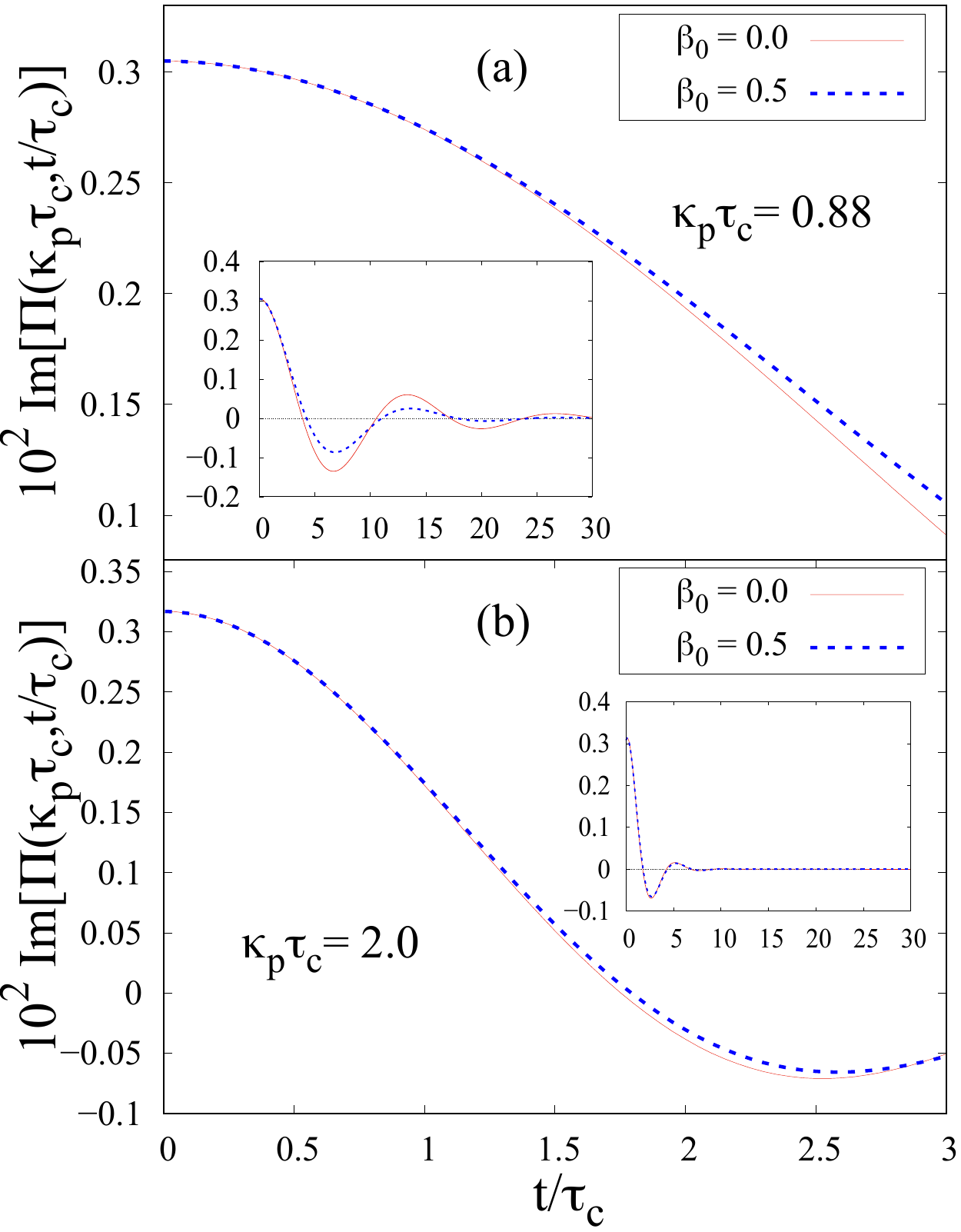}
 \caption{Time evolution of the most dominant mode of energy
 density fluctuations (a) for $\kappa_p \tau_c=0.88$ (corresponding
 to fluctuation parameter $r=2.0$) and (b) for $\kappa_p \tau_c=2.0$
 (for $r=4.8$). The results are in the absence (solid line) and
 presence (dashed line) of magnetic field. Insets show the mode
 evolution for a longer time.}
 \label{fig5}
\end{figure}

For further analysis we focus on the time evolution of 
the most dominant mode of energy density fluctuations.
Figure \ref{fig5}(a) shows the time evolution of Im$[\Pi(
\kappa'_p,t')]$ for the dominant mode for $\kappa'_p =
\kappa_p \tau_c=0.88$ (corresponding to fluctuation parameter
$r=2.0$) in the absence (solid line) and presence (dashed line)
of magnetic field. It is clear that the effect of magnetic
field becomes noticeable at times $t/\tau_c \gtrsim 1$
that enforces a higher magnitude in the fluctuation strength
at early times followed by a gradual increase in the damping
of the oscillating modes (as shown in the inset). In Fig.
\ref{fig5}(b), we also show the the evolution of a high
momentum mode $\kappa_p \tau_c=2.0$ (for $r=4.8$). This high
momentum (short wavelength) mode is strongly suppressed and
quickly damped in presence of magnetic field.

The above results demonstrate that magnetic field can affect 
the evolution of energy density fluctuations in the transverse
plane at the timescale $t/\tau_c \sim 1$. Consequently, the
characteristics of fluctuations in three dimensional physical
space can get modified, which may generate additional spatial
anisotropies. For precise quantification of the growth of these
anisotropies due to magnetic field, it is necessary to perform
a full (6+1)-dimensional phase-space simulation. Nevertheless,
in the present (3+1)D evolution, the relation between the
dominant modes of fluctuations with and without magnetic field
can provide some crucial insight about this growth.

As mentioned above and evident from Figs. \ref{fig3} and
\ref{fig5}, the time evolution of ${\rm Im}[\Pi(\kappa'_p,t')]$
can be best fitted with the damped harmonic oscillator function:
\begin{equation}
 \text{Im}[\Pi(\kappa'_p,t')] = \alpha_0 \cos(\omega t' - \phi)
 \exp(-\gamma_m t').
\label{eq:dhf}
\end{equation}
Here $\alpha_0 \equiv \alpha_0(\kappa'_p)$ is the amplitude scale
factor of the oscillator, $\omega \equiv \omega(\kappa'_p)$ the
dimensionless angular frequency, $\phi \equiv \phi(\kappa'_p)$ the
phase, and $\gamma_m \equiv \gamma_m(\kappa'_p)$ the dimensionless
damping coefficient. We found that magnetic field has an
insignificant effect on the oscillatory factor $\alpha_0 \cos(
\omega t'-\phi)$, but increases the damping coefficient $\gamma_m$,
and further, $\omega$ is always greater than $\gamma_m$ 
\textemdash$~$representing an underdamped oscillator. This
yields a relation between the fluctuation modes with and without
$B$ as
\begin{equation}
\text{Im}[\Pi(\beta_0)] \approx \text{Im}[\Pi(0)] \exp[-\delta
\gamma_m \gamma_m(0) t'],
 \label{eq:scaling}
\end{equation}
where $\delta \gamma_m$=$[\gamma_m(\beta_0)$$-$$\gamma_m(0)]/
\gamma_m(0)$ is the fractional change in the damping coefficient
by the magnetic field, which is completely independent of the
initial magnitude of energy density fluctuations. The decay of
the modes in presence of $B$ can be then conveniently determined
by $\delta \gamma_m$.

\begin{figure}[t]
\includegraphics[width=0.99\linewidth]{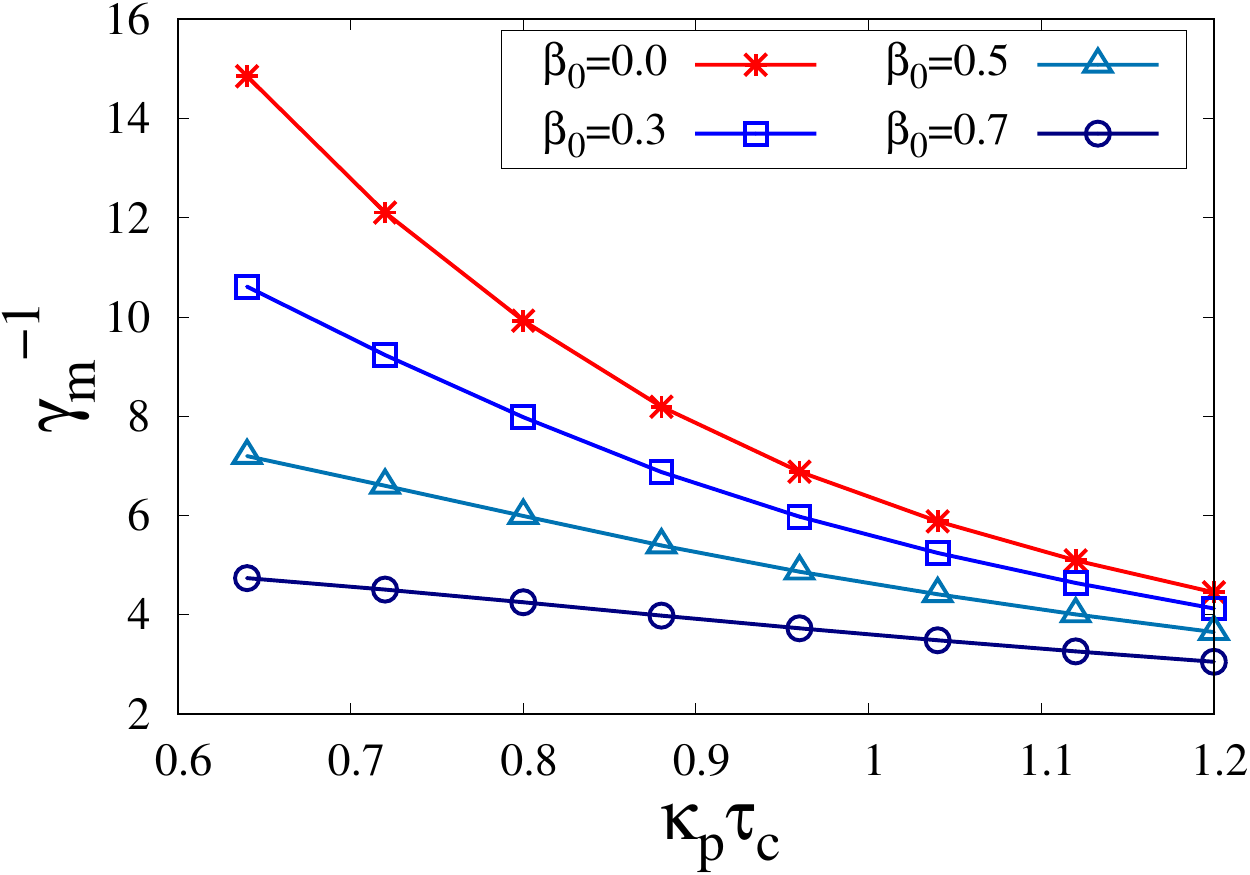}
 \caption{Variation of \textquotedblleft dimensionless decay
 timescale" $\gamma_m^{-1}$ of mode with peak-momentum $\kappa_p
 \tau_c$ for different values of magnetic parameter $\beta_0$.}
 \label{fig6}
\end{figure}

Note that $\gamma_m^{-1} \tau_c$ sets a decay timescale of the
mode, which can also be identified as the ``relaxation time" of
the particular mode. Figure \ref{fig6} shows the variation of the
\textquotedblleft dimensionless decay timescale" $\gamma_m^{-1}$
of mode with peak-momentum $\kappa_p\tau_c$ for different values
of magnetic field parameter $\beta_0$. The values of momentum in
the range $\kappa_p \tau_c \in [0.64,2.0]$ are obtained by varying
the fluctuation parameter between $r \in [0.5,4.8]$. In general,
for any value of $\beta_0$, the decay timescale $\gamma_m^{-1}$
exhibits a decreasing trend with increasing $\kappa_p \tau_c$ as
the higher momentum modes decay fast. Stronger magnetic field in
the medium, i.e. with increasing $\beta_0$, reduces the decay
timescale of especially the slow modes that actually experience
magnetic force for a longer duration \textemdash$~$inspite of
higher momentum modes are largely influenced (quantitatively) by
the magnetic field at any instant of time (see Fig. \ref{fig4}).

\begin{figure}[t]
\includegraphics[width=0.99\linewidth, angle=180]{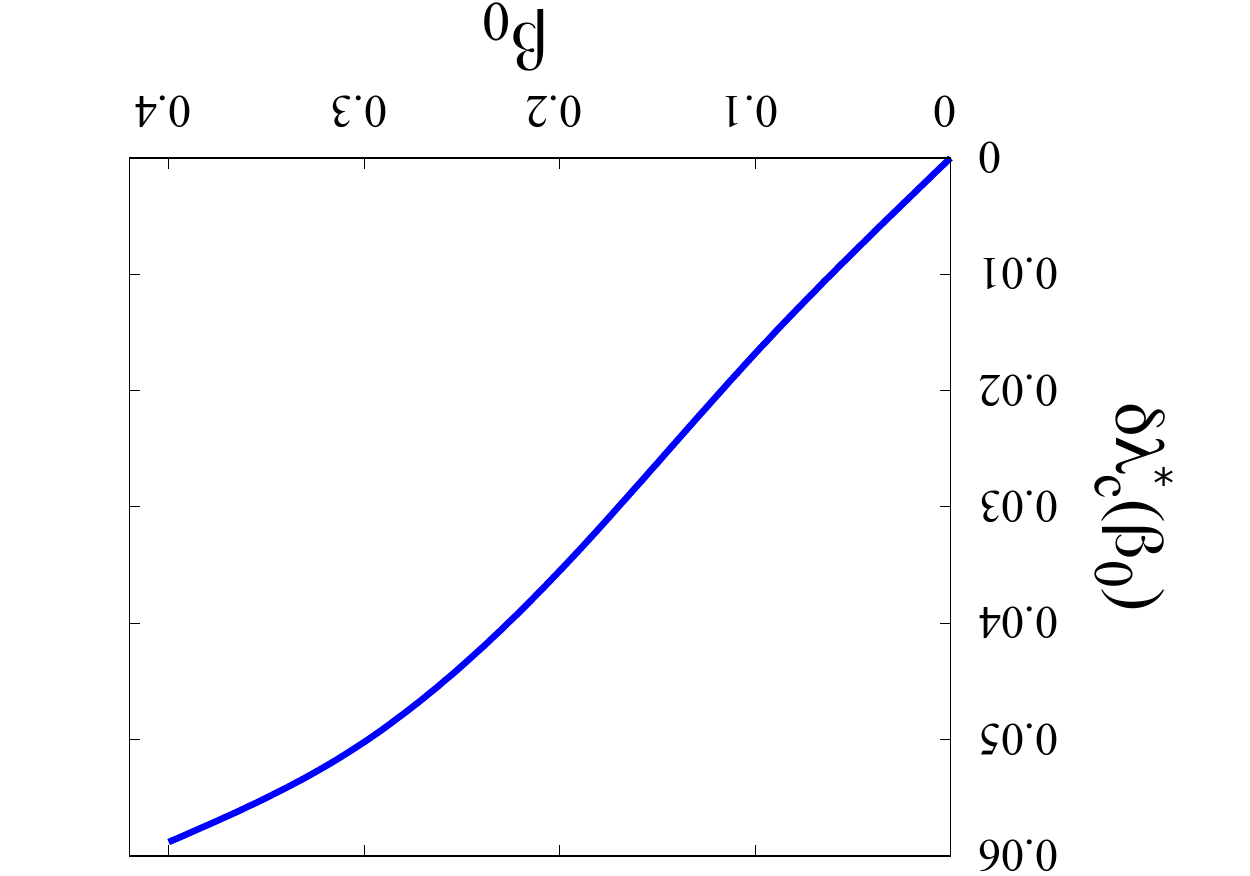}
 \caption{Magnetic field parameter $\beta_0$ dependence of the
 fractional change of wavelength cutoff $\delta
 \lambda^*_c(\beta_0)$=$[\lambda^*_c(\beta_0)- \lambda^*_c(0)]
 /\lambda^*_c(0)$. The bending in the curve at $\beta_0>0.3$ is
 due to the simulations and fitting procedures performed in a
 limited range of $\kappa_p \tau_c$.}
 \label{fig7}
\end{figure}

Each curve in Fig. \ref{fig6}, corresponding to a $\beta_0$ value,
can be best fitted with a power law scaling $\gamma_m^{-1}$=$
s_0(\kappa_p\tau_c)^{-s_1} $+$s_2$. The mode that has a decay
timescale comparable to the local-equilibration timescale of the
system can be identified with a momentum cutoff $\kappa_c \tau_c$
above which all the higher momentum modes are suppressed. In other
words, it resembles a \textquotedblleft dimensionless" wavelength
cutoff $\lambda^*_c=\lambda_c/\tau_c$ below which any inhomogeneity
in the energy density is suppressed. This cutoff can be determined
by putting $\gamma_m^{-1}=1$ in the above power law scaling. Figure
\ref{fig7} illustrates the qualitative growth of the wavelength
cutoff, $\lambda^*_c(\beta_0)$, with the magnetic field parameter
$\beta_0$; the value of $\lambda_c(\beta_0=0)$ turns out to be about
$2.5\tau_c$. Note that $\lambda_c (\beta_0)$ is equivalent to the
coarse-grained length scale inherent in the hydrodynamic description
for medium evolution. The above analysis indicates that magnetic
field suppresses the short wavelength fluctuations up to a larger
wavelength (as compared to $B=0$) resulting in a smoother
coarse-grained structure of the hydrodynamic variables.

Given that $\delta \gamma_m$ is enhanced more towards the lower
momentum modes (see Fig. \ref{fig6}), we can also obtain a power
law scaling behavior of $\delta \gamma_m$ with $\kappa_p \tau_c$
and $\beta_0$ (as found for $\gamma_m^{-1}$), namely:
\begin{equation}
 \delta \gamma_m \approx \beta_0^2 \left[
2 (\kappa_p \tau_c)^{-(0.03/\beta_0 +2)}-0.5 \right].
\label{eq:dgm}
\end{equation}
This relation is completely independent of the choice of initial
energy density fluctuations, and perfectly valid in the above
mentioned range of $\kappa_p \tau_c$. From the above relation one
can determine the fractional change in the modes, generated by the
magnetic field, by using the expression $\delta \Pi = ({\rm Im}[\Pi
(\beta_0)]$$-$${\rm Im}[\Pi(0)])/{\rm Im}[\Pi(0)]$. Although the
damping coefficient of modes and effects of magnetic field would
be quite different in three dimensional physical space, $\delta
\Pi$ may be valid in that case as well. Hence, using Eqs. 
\eqref{eq:scaling} and \eqref{eq:dgm}, $\delta \Pi$ provides a
measure of spatial anisotropies in the energy density solely
generated by the magnetic field.

\subsection*{B. Evolution of $\hat T^{\mu \nu}$ and dependence
on $\mu_Q$}

\begin{figure}[t]
\includegraphics[width=0.49\linewidth]{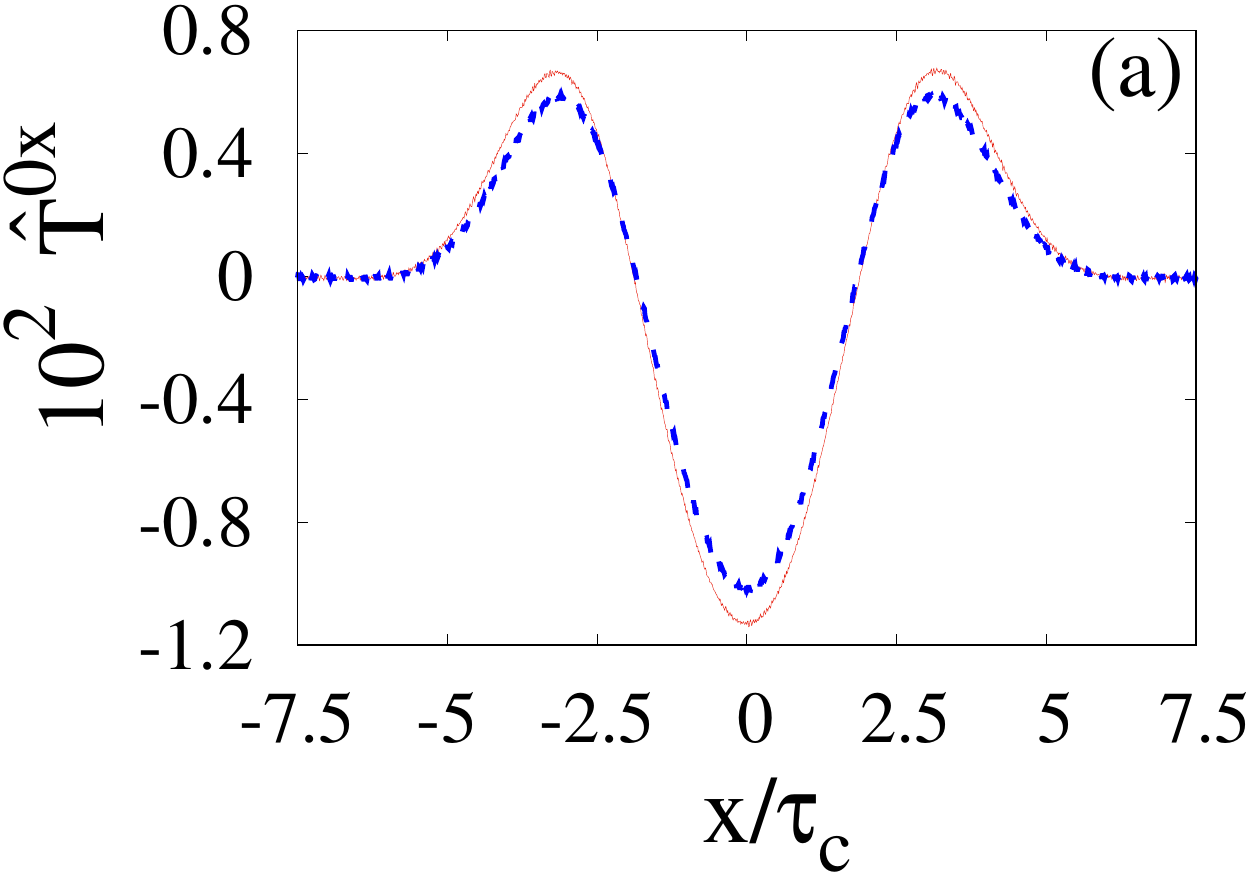}
\includegraphics[width=0.49\linewidth]{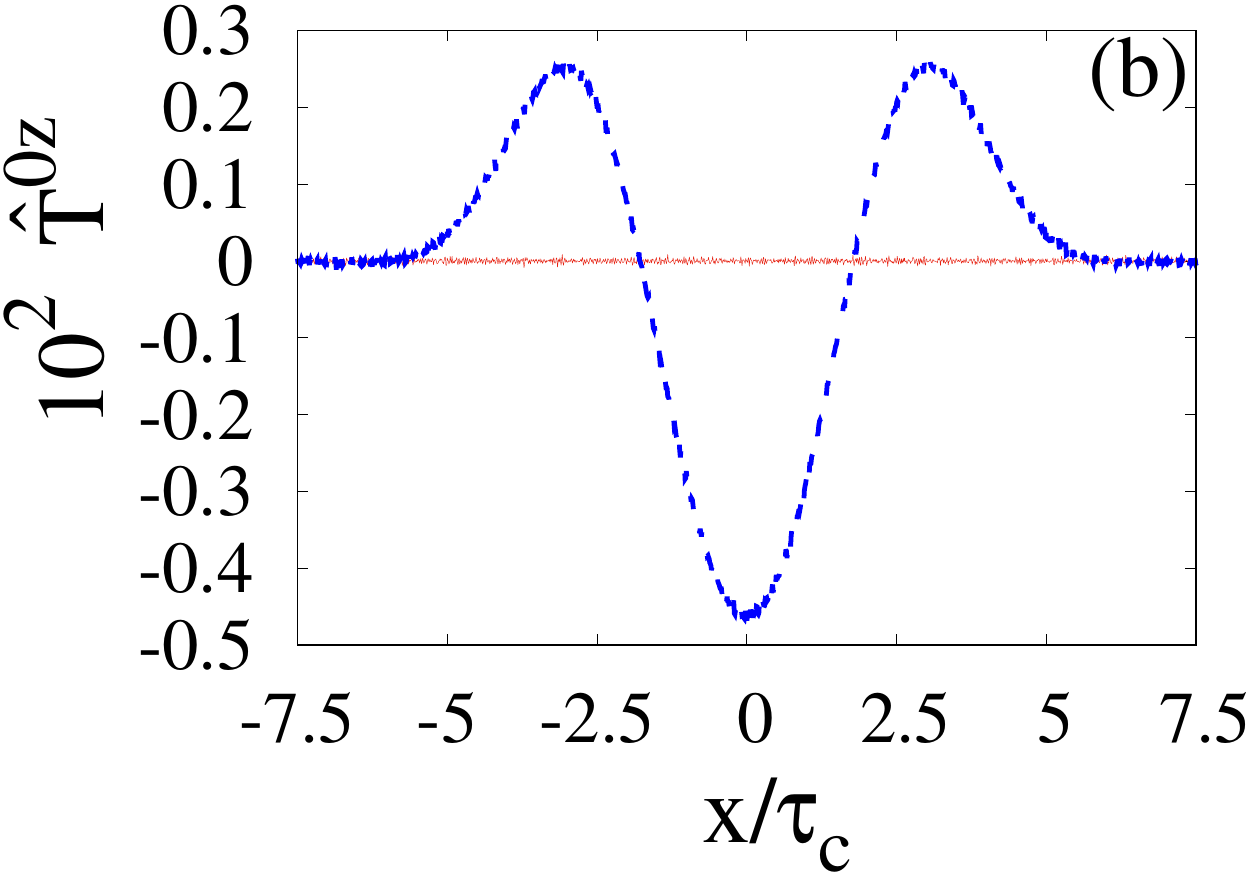}
\includegraphics[width=0.49\linewidth]{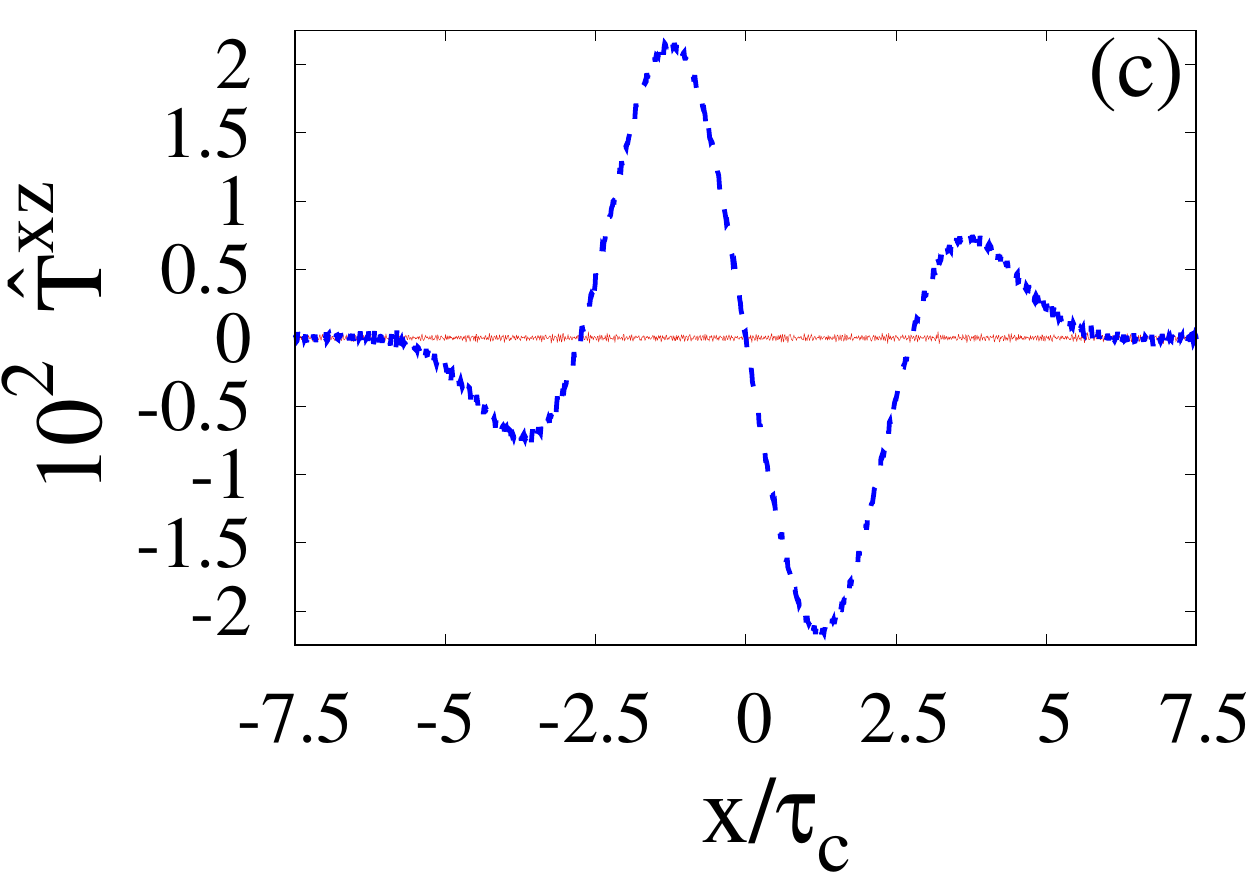}
\includegraphics[width=0.49\linewidth]{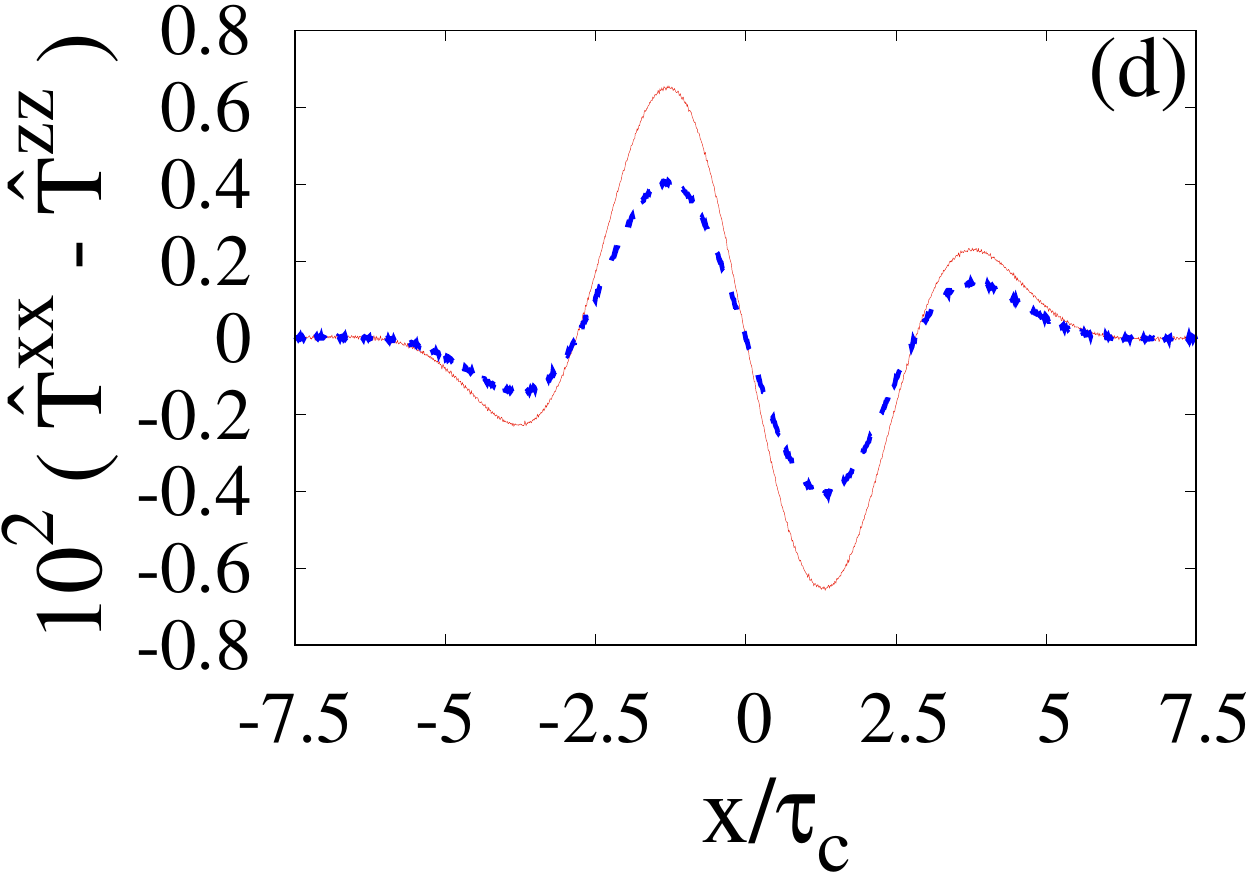}
 \caption{Spatial variation of the components of energy-momentum
 tensor (a) $\hat T^{0x}$
 (b) $\hat T^{0z}$ (c) $\hat T^{xz}$ and (d) ($\hat T^{xx} -
 \hat T^{zz}$)
 at a time $t/\tau_c=3.0$ in the absence (solid line) and
 presence (dashed line) of the magnetic field for the same simulation
 parameters as used in Fig. \ref{fig1}.}
 \label{fig8}
\end{figure}

In the previous subsection, the effects of magnetic field 
on the evolution of energy density fluctuations ($\delta
\hat\varepsilon =\delta \hat T^{00}$) have been studied and
found to increase the damping coefficient of mode oscillations.
In this subsection, we shall explore magnetic effects on the
other components of energy-momentum tensor $\hat T^{\mu \nu}$
which is calculated by using Eq. \eqref{eq:tmn} and performing
the momentum integrals as done in Eq. \eqref{eq:d2slv}. The
spatial variation of the energy-momentum tensor components,
$\hat T^{0x}$, $\hat T^{0z}$, $\hat T^{xz}$, and ($\hat T^{xx}
-\hat T^{zz}$), is shown in Fig. \ref{fig8} at time $t/\tau_c
=3.0$ in the absence (solid line) and in the presence (dashed
line) of magnetic field, for the same simulation parameters as
used in Fig. \ref{fig1}. The variation of $\hat T^{0x}$ arises
due to the spatial gradients present in the energy density
fluctuations as shown in Fig. \ref{fig1}. It is clear from Fig.
\ref{fig8} that the magnetic field suppresses $\hat T^{0x}$,
which essentially leads to the increase in the damping coefficient
of the mode oscillations (as shown previously). The most
appreciable effects of magnetic field involve the $z$-components,
namely $\hat T^{0z}$ and $\hat T^{xz}$, which become rather large
compared to vanishingly small values for $B=0$. Such a behavior
can be traced to the Lorentz force exerted by the magnetic field
that acts along $z'$ direction on the medium evolving along $x'$.
The magnetic field is seen to also suppress substantially the
magnitude of ($\hat T^{xx}-\hat T^{zz}$).

It is important to comment on the effects of charge chemical
potential $\mu_Q$ on the characteristic change seen for the
components of $\hat T^{\mu\nu}$ and, in particular, on the energy
density fluctuations in the presence of $B$. At $\mu_Q=0$
(when the net-electric charge density $\hat n = \hat n_+ - \hat
n_- = 0$ due
to identical particle and anti-particle number densities), the
equal and opposite Lorentz force exerted by the magnetic field
along $z'$ direction on the positive and negatively charged
particles causes $\hat T^{0z}$ and $\hat T^{xz}$ to vanish throughout
their evolution. Only at {\it finite} $\mu_Q$, the magnitude
of $\hat T^{0z}$ and $\hat T^{xz}$ become non-zero in presence of
magnetic field due to net-electric charge density imbalance
(as seen in Fig. \ref{fig8}). This also suggests, that
irrespective of the value of $\mu_Q$, a finite $B$ will affect
solely the magnitude of the momentum density $\hat T^{0x}$ and its
accompanied energy density fluctuations $\delta \hat\varepsilon$.
Consequently, the earlier discussed scaling of $\delta
\gamma_m$ for a given $\beta_0$  will remain unaltered which
we have verified for a range of $\mu_Q/T_0 = 0 - 2.0$ at a
fixed equilibrium temperature $T_0$.

\subsection*{C. Fluctuations with multiple modes}

\begin{figure}[t]
\includegraphics[width=0.99\linewidth]{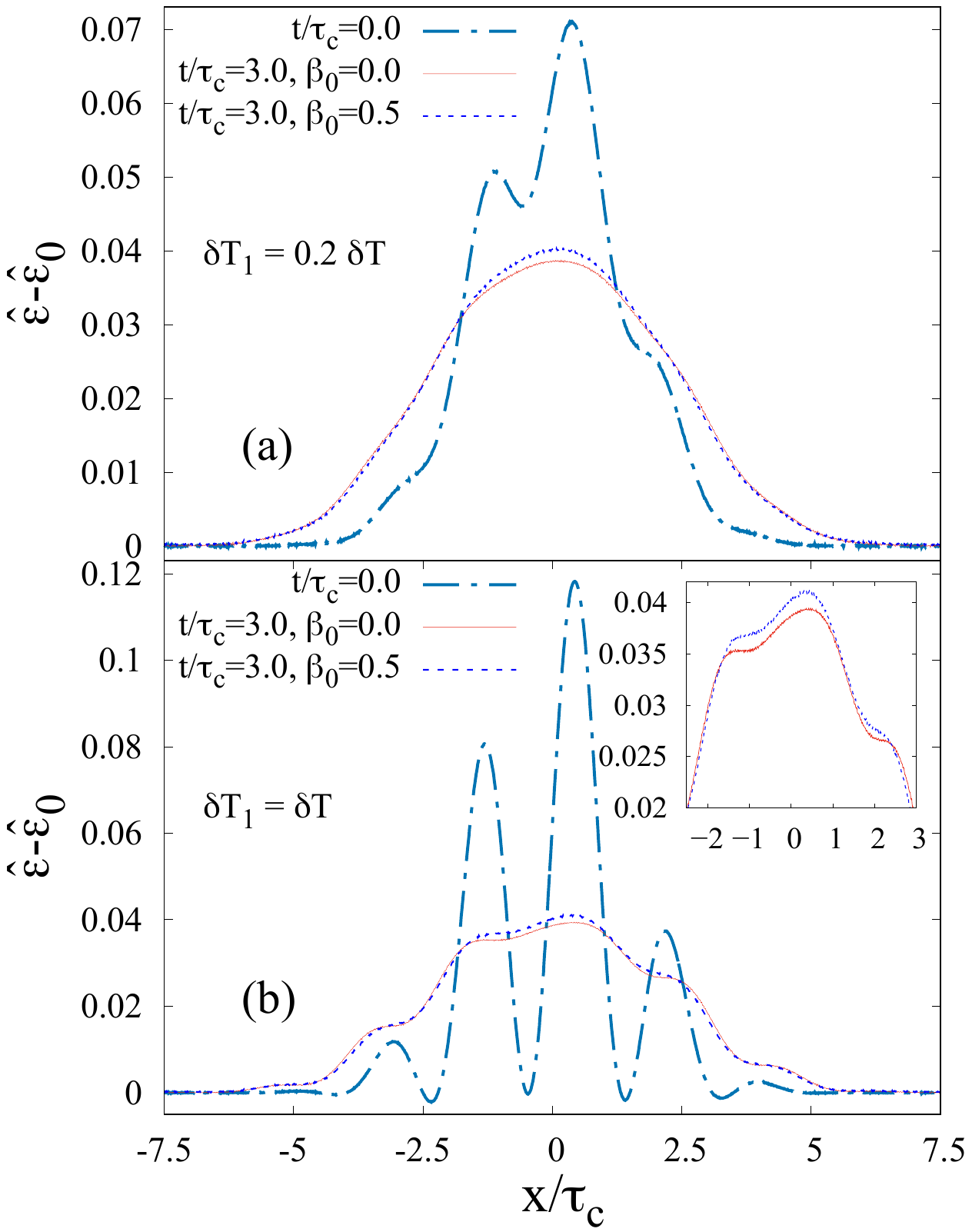}
 \caption{The energy density fluctuations, $(\hat\varepsilon-
 \hat\varepsilon_0)$, at the initial time (dash-dotted lines) and
 at time $t/\tau_c=3.0$ in the absence (solid lines) and presence
 (dashed lines) of the magnetic field with $\beta_0=0.5$. The
 result are for mode mixing parameter (a) $\delta T_1=0.2\delta
 T$ and (b) $\delta T_1 = \delta T$, where $\delta T = 0.01
 T_0$. It is clear from (a) and inset of (b) that in presence of
 $B$, the energy density profile is smoother and has larger peak
 value as compared to $B=0$.}
 \label{fig9}
\end{figure}

To illustrate the effects of magnetic field on the energy
density fluctuations having more than one initial dominant
mode, we consider fluctuations of the form
\begin{equation}
\begin{split}
 T(x')=T_0 + \Big[ \delta T \cos\Big(\frac{2\pi r_1 x'}{L}
 \Big) + \delta T_1 \sin\Big(\frac{2\pi r_2 x'}{L}\Big)
 \Big] \\
 \times \exp (-x'^2/2 \chi^2),~~~~~~~~~~~~~~~~~~~~~~~~~~~~~
 \end{split}
 \label{eq:fluct2}
\end{equation}
where the mode mixing parameter $\delta T_1$ is varied between
zero and $\delta T$ to generate different energy density
fluctuations of different amplitudes. For the present study,
we have taken $r_1=0.5$ and $r_2=8.0$ which allows the
generation of a low and a very high momentum mode, respectively.
The choice of sine and cosine functions invokes some
arbitrariness (such as an asymmetry about $x'=0$) resembling to
some extent a general form for energy density fluctuations
present in a physical system, for example, the initial-state
fluctuations in the Glauber model \cite{Loizides:2017ack} or in
the gluon saturation model \cite{cgc} as commonly employed for
initial conditions in the modeling of relativistic heavy-ion
collisions. The other simulation parameters are the same as
taken previously. Figure \ref{fig9} shows spatial dependence for
the evolution of the energy density fluctuations,
$(\hat\varepsilon-\hat\varepsilon_0)$. The results are for a
relatively small mode mixing $\delta T_1 = 0.2 \delta T$ as shown
in Fig. \ref{fig9}(a) and for maximal mixing $\delta T_1 = \delta
T$ as shown in Fig. \ref{fig9}(b). The initial energy density
profile (dash-dotted lines) becomes rather smooth at a later time
of $t/\tau = 3.0$ (solid and dashed lines). This arises as the
fluctuations of wavelengths shorter than the cutoff $\lambda^*_c$
are dominantly suppressed at times $t/\tau_c \gtrsim 1$. In
presence of magnetic field, at $t/\tau = 3.0$ (dashed lines), the
energy density profile becomes slightly smoother and has a larger
peak value as compared to $B = 0$ (solid lines); see inset of Fig.
\ref{fig9}(b). This smoothening is essentially caused by the
enhanced suppression of the short wavelength (high momentum) modes,
whereas the larger peak is due to slow dissipation of long wavelength
(low momentum) fluctuations at early times (as discussed in Fig.
\ref{fig5}).

\begin{figure}[t]
\includegraphics[width=0.99\linewidth]{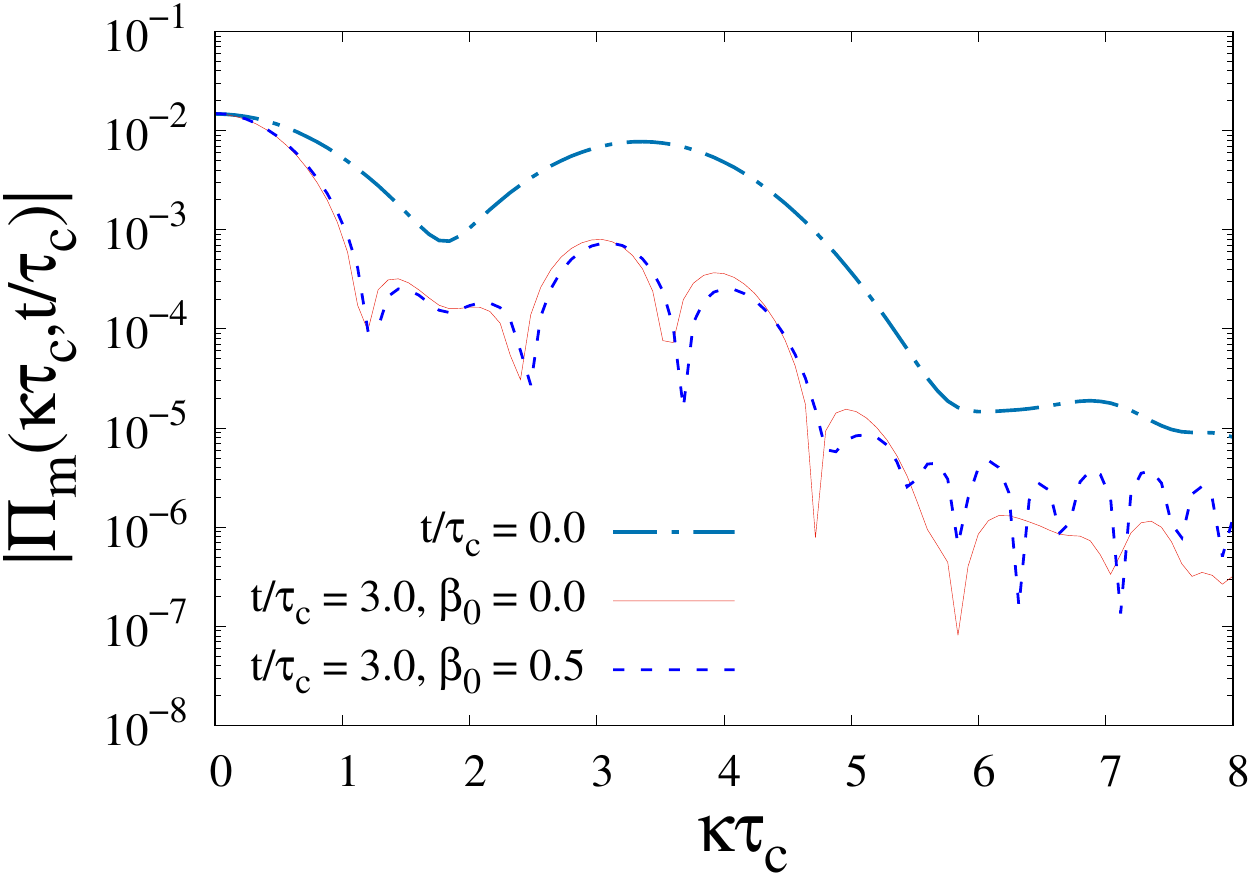}
 \caption{Momentum spectrum of energy density fluctuations
 at initial time (dash-dotted line), and at time $t/\tau_c=
 3.0$ in the absence (solid line) and in the presence
 (dashed line) of magnetic field for $\delta T_1=\delta T$.}
 \label{fig10}
\end{figure}

We present in Fig. \ref{fig10} the momentum spectrum [the
modulus of Fourier modes, $|\Pi_m(\kappa',t')|$, versus
$\kappa \tau_c$] of the energy density fluctuations of
Fig. \ref{fig9}(b). In presence of magnetic field (dashed
line), the high momentum dominant modes (at $\kappa \tau_c
\approx $ 3.0 and 4.0) are more suppressed, while the low
momentum modes ($\kappa \tau_c < 1.2$) are somewhat less
dissipated [as compared to $B=0$ (solid line)] at later time
$t/\tau_c = 3.0$. All these lead to qualitatively different
characteristics of the fluctuations in the presence of
magnetic field.

\section*{V. Phenomenological Implications}

We shall discuss the phenomenological implications of the 
present study on certain important features and its related
observables pertaining to relativistic heavy and light ion
collisions, although it is potentially applicable to any small
system whose constituents are electrically charged. For Au+Au
collisions at c.m. energy $\sqrt{s_{NN}} =200$ GeV and impact
parameter $b=8.0$ fm, the magnetic field can be as large as $|qB|
\simeq 0.1 m_\pi^2$ \cite{BMEq} at a proper time $\tau \sim 0.2$
fm/c which is the timescale for saturated gluonic configuration
to decay into (anti-)quarks and gluons \cite{shuryak0,BMEq}. In
the gluon saturation model, thermalization or hydrodynamization
occurs at $\tau_{\rm eq} \simeq 0.4$ fm \cite{hydroSim} when the
magnetic field can still survive with appreciable strength; the
magnetic field may persist in the entire partonic and possibly
hadronic phase if the medium has a large electrical conductivity
\cite{Hattori:2016lqx}.

In the pre-equilibration dynamics at $\tau \lesssim 0.4$ fm,
various short wavelength modes are present in the system with
decay timescales smaller or comparable to this time (see Fig. 2
in Ref. \cite{Schenke12}). Such fluctuation modes have sources
from initial-state fluctuations in nucleon position, parton
production and dynamics, hadron production and evolution. The
magnetic field can increase the damping coefficient of mode
oscillations and modify the characteristics of short wavelength
fluctuations in the reaction plane (transverse to $B$) as
demonstrated in this work. Consequently, this can have measurable
effects on the azimuthal anisotropy of particle production/emission,
namely the collective flow harmonics $v_n(p_T) = \langle \cos n(\phi-
\Psi_n)\rangle$ (especially the odd harmonics that are driven by
initial-state fluctuations) and the flow fluctuations
\cite{Bhalerao:2014xra,Giacalone:2017uqx,saumia1,Schenke12i,saumia2,
ourRev,ehr}. In particular, the flow and flow-fluctuation observables
would exhibit a noticeable suppression, reflecting the enhanced
damping of fluctuation modes over the evolution time as found here
(Fig. \ref{fig6}). Moreover, a smoother energy density profile,
induced by suppression of short wavelength fluctuations in presence
of magnetic field, should also show some qualitative changes in the
power spectrum of flow fluctuations [$v_n(p_T)$ versus $n$] at higher
$n$.

The hydrodynamic fluctuations \cite{kapusta1,spal1,
Chattopadhyay:2018dth} and disturbances in the medium due to
energy deposition by a partonic jet 
\cite{Casalderrey-Solana:2016jvj,Tachibana:2017syd,
Milhano:2017nzm} can prevail during the entire evolution of
the system. The thermal or hydrodynamic fluctuations are
correlated over short length scales that generate short-range
correlation peak at small rapidity separation $\Delta y = y_1
-y_2$ and nontrivial structure at large $\Delta y$ in the
two-particle rapidity correlation \cite{kapusta1,spal1,
Chattopadhyay:2018dth}. Such long-range rapidity structures
have been observed in multiparticle correlation measurements
involving heavy ion and high-multiplicity light particle
collision experiments at relativistic energies. Our analysis
suggests magnetic damping of the peak at $\Delta y \approx 0$
and farther spread of the correlations in the rapidity
separation. The disturbance generated by energy-momentum
deposited in the vicinity of traversing hard jets in QGP and
modifications of the jet shape and jet substructure 
observables due to (enhanced) rescattering of the emitted soft
gluons in the medium will be sensitive to the magnetic field.

On the other hand, near the critical end point in the QCD
phase diagram, the correlations among fluctuations diverge
resulting in new fluctuation modes \cite{Stephanov:2010zz,
An:2019csj}. The non-monotonous behavior in the event-by-event
fluctuations with varying c.m. energy $\sqrt{s_{NN}}$ signals
the location of the critical point. As the QCD critical point
is expected to be at finite baryon density and at moderate
collision energy \cite{Pandav:2022xxx}, the strength of magnetic
field will be relatively smaller, which however, decays slowly
and can slow down some of the modes of critical fluctuations
via increasing the damping coefficient. This can essentially
enhance the magnitude of the observable signatures of the
critical point. A detailed numerical simulation involving all
the discussed features can provide quantitative effects of the
magnetic field.

\section*{VI. Summary and Conclusions}

In this paper we have studied the effects of magnetic field on
the evolution of energy density fluctuations in the transverse
direction. We find characteristic changes in the fluctuations
at the timescale required by the system to achieve local thermal
equilibrium. The magnetic damping of the underlying medium slows
down the dissipation of the initial strength of the fluctuation
near its peak while spreading out the fluctuation spatially to
larger distances at early times. Increased Lorentz force at
later times enforces larger damping of the fluctuation compared
to the field-free case. A detailed Fourier mode analysis of the
energy density fluctuations reveals that the low momentum modes,
which survive longer in the entire evolution of the system, are
strongly damped as compared to the fast evolving high momentum
modes. The behavior is found to progressively increase with the
strength of the magnetic field. However, at any instant of time,
the magnetic field affects quantitatively more the high momentum
modes as compared to the low momentum modes. This leads to a
growth in the cutoff for the shortest wavelength fluctuations
present in the system. If this cutoff is identified with the
coarse-grained length scale in the hydrodynamic description of
the medium, which then indicates an enhanced smoothening of
energy density profile in presence of magnetic field. Further,
the fluctuations in the direction transverse to the magnetic field
are essentially affected, and moreover, various components of the
energy-momentum tensor in the transverse direction are found to be
modified differently, namely, $\hat T^{0x}$ is suppressed, while
the $z$-components $\hat T^{0z}$ and $\hat T^{xz}$ are generated
for fluid evolving along $x$-direction. As a result, additional
spatial and momentum anisotropies can be generated in the three
dimensional physical space by the magnetic field.

The present study has crucial phenomenological implications
in the context of understanding the properties of quark-gluon
plasma formed in relativistic heavy-ion collisions, and in
general for any small systems whose constituents are electrically
charged. In particular, the magnetic field can have noticeable
impact on the potentially important observables, as for example,
the flow harmonics and flow fluctuations, the hydrodynamic
fluctuations, the jet substructure, and the dynamics of
correlations and fluctuations close to the QCD critical point.

In the study, we have worked in the relaxation time approximation
where the nonlinear effects, which can arise due to proper
collision integral, has been ignored. The  non-linearity can
affect the evolution of energy density fluctuations as well as
the effects of magnetic field studied in this work. We defer its
inclusion for future studies.

\section{acknowledgments}
We would like to thank Rajeev Bhalerao and Sunil
Jaiswal for useful discussions. The simulations
were performed at the High Performance Computing
cluster at TIFR, Mumbai. The authors acknowledge
financial support by the Department of Atomic
Energy (Government of India) under Project
Identification No. RTI 4002.


\begin{thebibliography}{99}

\bibitem{cmbr_mag} J. Adams, U.H. Danielsson, D. Grasso, and 
H. Rubinstein, Phys. Lett. B {\bf 388}, 253 (1996).

\bibitem{cllpsmag} H. Sotani, Phys. Rev. D {\bf 79}, 084037 (2009).

\bibitem{BNSmag0} K. Kiuchi, K. Kyutoku, Y. Sekiguchi, M. Shibata,
and T. Wada, Phys. Rev. D {\bf 90}, 041502(R) (2014).

\bibitem{BNSmag1} K. Kiuchi, P. Cerd\'{a}-Dur\'{a}n, K. Kyutoku,
Y. Sekiguchi, and M. Shibata, Phys. Rev. D {\bf 92}, 124034 (2015).

\bibitem{BNSmag2} K. Dionysopoulou, D. Alic, and L. Rezzolla, 
Phys. Rev. D {\bf 92}, 084064 (2015).

\bibitem{BNSmag3} A. Harutyunyan, A. Nathanail, L. Rezzolla, and A. 
Sedrakian, Eur. Phys. J. A {\bf 54}, 191 (2018). 

\bibitem{BNSmag4} K. Kiuchi, K. Kyutoku, Y. Sekiguchi, and M. 
Shibata, Phys. Rev. D {\bf 97}, 124039 (2018).

\bibitem{BNSmag5} M. Anderson {\it et al.}, Phys. Rev. Lett. {\bf 
100}, 191101 (2008).

\bibitem{BNSmag6} T. Kawamura {\it et al.}, Phys. Rev. D {\bf 94},
064012 (2016).

\bibitem{BNSmag7} A. Endrizzi, R. Ciolfi, B. Giacomazzo, W. Kastaun,
and T. Kawamura, Class. Quantum Grav. {\bf 33}, 164001 (2016).

\bibitem{BNSmag8} R. Ciolfi, Gen. Relativ. Gravit. {\bf 52}, 59 (2020). 

\bibitem{larry} D.E. Kharzeev, L.D. McLerran, and H.J. Warringa, 
Nucl. Phys. A {\bf 803}, 227 (2008).

\bibitem{lateTmB} B. M${\rm \ddot{u}}$ller and A. Sch${\rm \ddot{a}}
$fer, Phys. Rev. D {\bf 98}, 071902(R) (2018).

\bibitem{tuchin0} K. Tuchin, J. Phys. G: Nucl. Part. Phys. 
{\bf 39}, 025010 (2012).

\bibitem{KharzMHD} U. G${\rm \ddot{u}}$rsoy, D. Kharzeev, and K.
Rajagopal, Phys. Rev. C {\bf 89}, 054905 (2014).

\bibitem{ranjita} R.K. Mohapatra, P.S. Saumia, and A.M. Srivastava, 
Mod. Phys. Lett. A {\bf 26}, 2477 (2011).

\bibitem{ourMHD} A. Das, S.S. Dave, P.S. Saumia, and A.M. Srivastava,
Phys. Rev. C {\bf 96}, 034902 (2017).

\bibitem{ourRev} S.S. Dave, P.S. Saumia, and A.M. Srivastava, 
Eur. Phys. J. Spec. Top. {\bf 230}, 673 (2021).

\bibitem{mhd1} G. Inghirami {\it et al.}, Eur. Phys. J. C
{\bf 76}, 659 (2016).

\bibitem{mhd2} G. Inghirami {\it et al.}, Eur. Phys. J. C
{\bf 80}, 293 (2020).

\bibitem{Landau_em} L.D. Landau and E.M. Lifshitz, {\it
Electrodynamics of Continuous Media, Vol.8}, Second Edition 
(Pergamon Press Ltd., 1984).

\bibitem{magInfl1} T. Kobayashi, J. Cosmol. Astropart. Phys.
05 (2014) 040.

\bibitem{magInfl2} A. Talebian, A. Nassiri-Rad, and H. Firouzjahi,
Phys. Rev. D {\bf 105}, 103516 (2022).

\bibitem{RevQGP1} S. Schlichting and D. Teaney, Annu. Rev. Nucl. 
Part. Sci. {\bf 69}, 447 (2019). 

\bibitem{RevQGP2} J. Berges, M.P. Heller, A. Mazeliauskas, 
and R. Venugopalan, Rev. Mod. Phys. {\bf 93}, 035003 (2021).

\bibitem{RevQGP3} H. Elfner and B. M\"{u}ller, 
arXiv:2210.12056 [nucl-th].

\bibitem{tuchin1} K. Tuchin, Phys. Rev. C {\bf 88}, 024911
(2013).

\bibitem{tuchin2} E. Stewart and K. Tuchin, Phys. Rev. C 
{\bf 97}, 044906 (2018).

\bibitem{BMEq} J.-J. Zhang {\it et al.}, Phys. Rev. Research 
{\bf 4}, 033138 (2022).

\bibitem{hydroSim} B. Schenke, S. Jeon, and C. Gale, Phys. Rev.
Lett. {\bf 106}, 042301 (2011).

\bibitem{Schenke12} B. Schenke, P. Tribedy, and R.
Venugopalan, Phys. Rev. Lett. {\bf 108}, 252301 (2012).

\bibitem{kapusta1} J.I. Kapusta, B. M\"{u}ller, and M. Stephanov,
Phys. Rev. C {\bf 85}, 054906 (2012).

\bibitem{spal1} C. Chattopadhyay, R.S. Bhalerao, and S. Pal,
Phys. Rev. C {\bf 97}, 054902 (2018).

\bibitem{stephanov1} X. An, G. Ba\c{s}ar, M. Stephanov, and H.-U. 
Yee, Phys. Rev. C {\bf 100}, 024910 (2019).

\bibitem{stephanov2} M. Stephanov and Y. Yin, Phys. Rev. D {\bf
98}, 036006 (2018).

\bibitem{saumia1} A.P. Mishra, R.K. Mohapatra, P.S. Saumia, and 
A.M. Srivastava, Phys. Rev. C {\bf 77}, 064902 (2008); Phys. Rev. 
C {\bf 81}, 034903 (2010).

\bibitem{Schenke12i} B. Schenke, S. Jeon, and C. Gale, Phys. 
Rev. C {\bf 85}, 024901 (2012).

\bibitem{saumia2} P.S. Saumia and A.M. Srivastava, Mod. Phys. Lett. 
A {\bf 31}, 1650197 (2016).

\bibitem{ehr} W. Ke and Y. Yin, arXiv:2208.01046 [nucl-th].

\bibitem{rischke1} G.S. Denicol, X.-G. Huang, E. Moln\'ar, 
G.M. Monteiro, H. Niemi, J. Noronha, D.H. Rischke, and Q. Wang,
Phys. Rev. D {\bf 98}, 076009 (2018).

\bibitem{rischke2} G.S. Denicol, E. Moln\'ar, H. Niemi, and 
D.H. Rischke, Phys. Rev. D {\bf 99}, 056017 (2019).

\bibitem{Ashu1} A.K. Panda, A. Dash, R. Biswas, and V. Roy,
J. High Energy Phys. 03 (2021) 216; Phys. Rev. D 
{\bf 104}, 054004 (2021).

\bibitem{bgk} P.L. Bhatnagar, E.P. Gross, and M. Krook, 
Phys. Rev. {\bf 94}, 511 (1954).

\bibitem{Anderson} J.L. Anderson and H.R. Witting, Physica 
{\bf 74}, 466 (1974).

\bibitem{PaulR} P. Romatschke, Eur. Phys. J. C {\bf 76}, 352 (2016).

\bibitem{heinz1} D. Bazow, M. Martinez, and U. Heinz, Phys. Rev. D {\bf 
93}, 034002 (2016).

\bibitem{kurkela1} M.P. Heller, A. Kurkela, M. Spali\'{n}ski, and V. 
Svensson, Phys. Rev. D {\bf 97}, 091503(R) (2018).

\bibitem{PaulR0} J. Brewer and P. Romatschke, Phys. Rev. Lett. {\bf 
115}, 190404 (2015).

\bibitem{PaulR1} P. Romatschke, Eur. Phys. J. C {\bf 77}, 21 (2017).

\bibitem{kurkela2} A. Kurkela, U.A. Wiedemann, Eur. Phys. J. 
C {\bf 79}, 776 (2019).

\bibitem{groot} S.R. de Groot, W.A. van Leeuwen, Ch.G. van Weert, 
{\it Relativistic Kinetic Theory: Principles and Applications}, 
(North-Holland Publishing Company, Amsterdam, 1980).

\bibitem{bamps1} M. Greif, C. Greiner, and Z. Xu, Phys. Rev. C 
{\bf 96}, 014903 (2017).

\bibitem{transMag1} Z. Chen, C. Greiner, A. Huang, and Z. Xu, 
Phys. Rev. D {\bf 101}, 056020 (2020).

\bibitem{transMag2} M. Kurian and V. Chandra, Phys. Rev. D {\bf 97},
116008 (2018).

\bibitem{sunil2022} D. Dash, S. Bhadury, S. Jaiswal, and A. Jaiswal,
Phys. Lett. B {\bf 831}, 137202 (2022).

\bibitem{landau_k} L.D. Landau and E.M. Lifshitz, {\it Course of
Theoretical Physics, Vol.10: Physical Kinetics} (Pergamon Press Ltd.,
Great Britain, 1981).

\bibitem{spal2} S. Jaiswal, C. Chattopadhyay, L. Du, U.
Heinz, and S. Pal, Phys. Rev. C {\bf 105}, 024911 (2022).

\bibitem{chandro22} C. Chattopadhyay, U. Heinz, and T. Sch$
{\rm \ddot{a}}$fer, Phys. Rev. C {\bf 107}, 044905 (2023).

\bibitem{bgk1} C. Manuel and S. Mr\'owczy\'nski, Phys. Rev. D {\bf 70},
094019 (2004).

\bibitem{bgk2} B. Schenke, M. Strickland, C. Greiner, and M.H. Thoma,
Phys. Rev. D {\bf 73}, 125004 (2006).

\bibitem{rafelski1} M. Formanek, C. Grayson, J. Rafelski, and
B. M${\rm \ddot{u}}$ller, Annals Phys. {\bf 434}, 168605 (2021).

\bibitem{rafelski2} C. Grayson, M. Formanek, J. Rafelski, and B.
M${\rm \ddot{u}}$ller, Phys. Rev. D {\bf 106}, 014011 (2022).

\bibitem{pracheta} P. Singha, S. Bhadury, A. Mukherjee, and
A. Jaiswal, arXiv:2301.00544 [nucl-th].

\bibitem{novelRTA} G.S. Rocha, G.S. Denicol, and J. Noronha, Phys.
Rev. Lett. {\bf 127}, 042301 (2021).

\bibitem{hicFluct} N. Borghini, M. Borrell, N. Feld, H. Roch, S. 
Schlichting, and C. Werthmann, Phys. Rev. C {\bf 107}, 034905 (2023).

\bibitem{leapfrog} W.H. Press, S.A. Teukolsky, W.T. Vetterling, and
B.P. Flannery, {\it Numerical Recipes in Fortran 77; The Art of
Scientific Computing}, 2nd ed. Vol. 1 (Syndicate, New York and
Melbourne, 1997).

\bibitem{rlx1} M. Prakash, M. Prakash, R. Venugopalan, and G. Welke,
Phys. Rep. {\bf 227}, 321 (1993).

\bibitem{rlx2} P. Kalikotay, S. Ghosh, N. Chaudhuri, P. Roy, and
S. Sarkar, Phys. Rev. D {\bf 102}, 076007 (2020).

\bibitem{star2} J. Adam {\it et al.} (STAR Collaboration), Phys.
Rev. C {\bf 98}, 014910 (2018).

\bibitem{star1} L. Adamczyk {\it et al.} (STAR Collaboration),
Nature (London) {\bf 548}, 62 (2017).

\bibitem{kapusLambda} L.P. Csernai, J.I. Kapusta, and T. Welle,
Phys. Rev. C {\bf 99}, 021901(R) (2019).

\bibitem{Loizides:2017ack}
C.~Loizides, J.~Kamin and D.~d'Enterria,
Phys. Rev. C \textbf{97}, 054910 (2018)
[erratum: Phys. Rev. C \textbf{99}, 019901 (2019)].

\bibitem{cgc} L. McLerran and R. Venugopalan, Phys. Rev. D 
{\bf 49}, 2233 (1994); Phys. Rev. D {\bf 49}, 3352 (1994).

\bibitem{shuryak0} E. Shuryak, Phys. Rev. C {\bf 80}, 054908 
(2009).

\bibitem{Hattori:2016lqx}
K.~Hattori, S.~Li, D.~Satow, and H.U.~Yee,
Phys. Rev. D \textbf{95}, 076008 (2017).

\bibitem{Bhalerao:2014xra}
R.S.~Bhalerao, J.Y.~Ollitrault, and S.~Pal,
Phys. Lett. B \textbf{742}, 94 (2015).

\bibitem{Giacalone:2017uqx}
G.~Giacalone, J.~Noronha-Hostler, and J.Y.~Ollitrault,
Phys. Rev. C \textbf{95}, 054910 (2017).

\bibitem{Chattopadhyay:2018dth}
C.~Chattopadhyay and S.~Pal,
Phys. Rev. C \textbf{98}, 034911 (2018).

\bibitem{Casalderrey-Solana:2016jvj}
J.~Casalderrey-Solana, D.~Gulhan, G.~Milhano, D.~Pablos, and K.~Rajagopal,
J. High Energy Phys. 03 (2017) 135.

\bibitem{Tachibana:2017syd}
Y.~Tachibana, N.B.~Chang, and G.Y.~Qin,
Phys. Rev. C \textbf{95}, 044909 (2017).

\bibitem{Milhano:2017nzm}
G.~Milhano, U.A.~Wiedemann, and K.C.~Zapp,
Phys. Lett. B \textbf{779}, 409 (2018).

\bibitem{Stephanov:2010zz}
M.A.~Stephanov,
Prog. Theor. Phys. Suppl. \textbf{186}, 434 (2010).

\bibitem{An:2019csj}
X.~An, G.~Ba\c{s}ar, M.~Stephanov, and H.U.~Yee,
Phys. Rev. C \textbf{102}, 034901 (2020).

\bibitem{Pandav:2022xxx}
A.~Pandav, D.~Mallick, and B.~Mohanty,
Prog. Part. Nucl. Phys. \textbf{125}, 103960 (2022).


\end{thebibliography}
\end{document}